\documentclass[twocolumn,superscriptaddress,preprintnumbers,showpacs,amsmath,amssymb,floatfix,endfloats,prb]{revtex4}

\usepackage{graphicx}
\font\bba=msbm10 scaled 1200
\font\bbb=msbm8 
\font\bbc=msbm6 
\newfam\bbfam
\textfont\bbfam=\bba
\scriptfont\bbfam=\bbb
\scriptscriptfont\bbfam=\bbc

\begin{document}

\title{Liquid-vapor transition of systems with mean field universality class}

\author{ Gernot J. Pauschenwein}
\email {pauschenwein@cmt.tuwien.ac.at} 
\affiliation{CMS and Institut f\"ur Theoretische Physik, TU Wien, \\
Wiedner Hauptstra{\ss}e 8-10, A-1040 Wien, Austria} 
\author{ Jean-Michel Caillol}
\email {jean-michel.caillol@th.u-psud.fr} 
\affiliation{Laboratoire de Physique Th\'{e}orique, UMR 8627\\
B\^{a}timent 210, Universit\'{e} Paris-Sud, 91405 Orsay Cedex, France} 
\author{ Dominique  Levesque}
\email {dominique.levesque@th.u-psud.fr} 
\affiliation{Laboratoire de Physique Th\'{e}orique, UMR 8627\\
B\^{a}timent 210, Universit\'{e} Paris-Sud, 91405 Orsay Cedex, France}
\author{ Jean-Jacques Weis}
\email {jean-jacques.weis@th.u-psud.fr}  
\affiliation{Laboratoire de Physique Th\'{e}orique, UMR 8627\\
B\^{a}timent 210, Universit\'{e} Paris-Sud, 91405 Orsay Cedex, France}
\author{ Elisabeth Sch\"oll-Paschinger}
\email {elisabeth.schoell-paschinger@univie.ac.at} 
\affiliation{CMS and Fakult\"at f\"ur Physik, Universit\"at Wien, \\
Boltzmanngasse 5, A-1090 Wien, Austria} 
\author{ Gerhard Kahl} 
\email {gkahl@tph.tuwien.ac.at} 
\affiliation{CMS and Institut f\"ur Theoretische Physik, TU Wien, \\
Wiedner Hauptstra{\ss}e 8-10, A-1040 Wien, Austria}

\date{\today}

\begin{abstract}
We have considered a system where the interaction, $v(r) = v_{\rm
IS}(r) + \xi^2 v_{\rm MF}(r)$, is given as a linear combination of two
potentials, each of which being characterized with a well-defined
critical behavior: for $v_{\rm IS}(r)$ we have chosen the potential
of the restricted primitive model which is known to belong to the
Ising 3D (IS) universality class, while for $v_{\rm MF}(r)$ we have
considered a long-range interaction in the Kac-limit, displaying
mean field (MF) behavior. We study the performance of two theoretical
approaches and of computer simulations in the critical region for this
particular system and give a detailed comparison between theories and
simulation of the critical
region and  the location of the critical point. Both, theory and
simulation give evidence that the system belongs to the MF
universality class for any positive value of $\xi$ and that it shows
only non-classical behavior for $\xi=0$. While in this limiting case
theoretical approaches are known to fail, we find good
agreement for the critical properties between the theoretical
approaches and the simulations for $\xi^2$ larger than 0.05.
\begin{center}\it
{Submitted to the Journal of Chemical Physics}
\end{center}
\end{abstract}


\maketitle

\section{Introduction}

The exact determination of the critical properties of a fluid (i.e., in terms
of the location of its critical point {\it and} of its universality class)
represents a formidable challenge both to theoretical approaches as well as to
computer simulations.  Although integral-equation theories are quite
successful in predicting the thermodynamic and structural properties of a
variety of simple and more complex systems over a wide domain in
temperature-density space \cite{Caccam:96} they meet with variable success in
the critical region due to an unsatisfactory treatment of long wavelength
fluctuations which are of particular relevance in this region
\cite{Parola:95,Parola:84}. To give an example, for the modified hypernetted
chain equation (MHNC), which is amongst the most accurate liquid state
theories \cite{HANSEN}, the boundary of the density region at which no
solution is found does not correspond to the spinodal and there is no
divergence of the compressibility. The mean spherical approximation (MSA), on
the other hand, satisfies scaling laws in the critical region, though with
peculiar critical exponents (mean spherical exponents \cite{Stell:69}), but,
similar to MHNC, fails to give a proper treatment of the first order
gas-liquid (GL) transition \cite{Parola:95}. For computer simulations, on the
other hand, suitable techniques have been developed that allow the
determination of the critical point via data extrapolation from simulations
performed in finite simulation volumes \cite{Wildin:92,Kim:04,Kim:03}.

The purpose of the present paper is to examine the performance of
theoretical approaches and of computer simulations in the critical
region for a system where the potential $v(r)$ is a linear
combination of two interactions, exhibiting different, but
well-established critical behavior, i.e., Ising 3D (IS) and
mean field (MF):
\begin{equation}
v(r) = v_{\rm IS}(r) + \xi^2  v_{\rm MF}(r).
\label{potential}
\end{equation}
The dimensionless parameter $\xi^2$ indicates the relative importance of
the two 
contributions. For reasons given below, we shall use a Coulombic
interaction for $v_{\rm IS}(r)$ and a Kac potential
\cite{Kac:59,Lebowi:66} for $v_{\rm
MF}(r)$ which we define as the limit, at fixed volume
$V$,
\begin{equation}
v_{\rm MF}(r) = \lim_{\alpha^* \to 0} \;  \alpha^{*3} \Phi(\alpha^* x)
\label{2} 
\end{equation}
where $\Phi(x)= - \displaystyle \frac{q^2}{\sigma} \frac{\mathrm e^{-x}}{x}$
($x=r/\sigma$, $\sigma$ length scale, $q$ has the dimension of a charge). When
the dimensionless parameter $\alpha^* =\alpha \sigma$ tends to zero the
strength of the potential decreases and its range increases to infinity. It is
crucial to observe that the limit $\alpha \to 0$ in Eq.\ (\ref{2}) must be
taken \textit{before} taking the thermodynamic limit, i.\ e.\, at fixed
volume. Although one might expect that the critical behavior of the system is
the result of a competition between the Ising 3D and the MF behavior, leading,
for example, to a cross over between the two universality classes, it turns
out that the criticality of the system is described by a MF behavior except at
$\xi^2 = 0$ where the system behaves Ising 3D like. This is unambiguously
confirmed both by simulations and theory (cf.\ Secs. III and IV).

To overcome the problem of liquid state theories mentioned above, two advanced
liquid state theories have been proposed in recent years that are able to
provide accurate data in the critical region: the hierarchical reference
theory (HRT) of Parola and Reatto \cite{Parola:84} and the self-consistent
Ornstein-Zernike approximation (SCOZA) originally proposed by H{\o}ye and
Stell \cite{Hoye:77,Hoye:84}. Compared to standard liquid state theories they
are more successful in describing the critical region, providing, in
particular, non-classical (i.e., non-MF) critical exponents. The HRT approach
is a successful merger \cite{Parola:95,Parola:84} of liquid state theory with
renormalization group ideas.  In the simplest form of SCOZA
\cite{MolPhys_95_483,MonatshChem_132_1413,EurophysLett_63_538} (which will be
used here) the Ornstein-Zernike (OZ) equation is supplemented by an
approximate, MSA-type closure relation for the direct correlation function
containing a density- and temperature-dependent function which is determined
by imposing self-consistency between the compressibility and internal energy
routes to the thermodynamics. SCOZA predicts non-MF critical exponents
\cite{Hoye:00}; they are different when approaching the critical temperature
from above or below and, in both cases, satisfy hyperscaling
\cite{Hoye:00}. We note that the formalism of SCOZA is not applicable for the
Kac potential (i.e., for $\alpha =0$) itself; instead, we rather have to
consider a potential with a small, but finite parameter $\alpha$. On the other
hand, this restriction has also the attractive feature that we can study in
detail how the critical behavior varies as $\alpha$ tends to zero; indeed we
can identify a cross-over behavior between a non-classical and a MF behavior.

In addition we use a recent improved mean field theory based on a loop
expansion of the free energy of the RPM \cite{Caillo:05,Caillo:06} to describe
the thermodynamic properties of our system.

Finally, we have used grand canonical Monte Carlo (GCMC)
simulations performed on a hypersphere \cite{Caillo:92,Caillo:93} together with histogram
reweighting techniques \cite{Swends:93}  and finite size scaling
analysis  to locate the
critical point and the near critical coexistence curve
\cite{Wildin:92,Wildin:95}.

With respect to the contribution to $v(r)$ with Ising 3D criticality
we resorted to the restricted primitive model (RPM). This choice has
three advantages: first, it is now well-established by simulations
that the RPM belongs to the Ising 3D universality class \cite{Caillo:02,Luijte:02}. In addition,
the critical parameters (i.e., temperature $T_c$ and density $\rho_c$)
and the near critical coexistence curve are accurately known. Second,
the reduced, dimensionless, critical temperature and density of the RPM $T_c
\approx 0.0489-0.0492$, $\rho_c \approx 0.076-0.080$
\cite{Caillo:02,Panagi:02,Luijte:02} (for the definition of the
reduced units see Sec.\ II) differ notably from those of the Kac
model ($\lim_{\xi^2\rightarrow0}T_c/\xi^2 \approx 1.13052$, $\rho_c \approx 0.27$) which
would not be the case by taking for $v_{\rm IS}(r)$ a
short range Yukawa potential. Third, the improved MF theory
outlined above can be formulated with closed expressions for a
size-symmetric (and possibly charge-asymmetric) system of charged hard
spheres, including thus the RPM \cite{Caillo:05}. These attractive features are
contrasted by one serious disadvantage: theoretical approaches that
have been proposed up to date in the literature are not able to
provide results for $T_c$ and $\rho_c$ for the RPM that are in
reasonable agreement with the  predictions of simulation
\cite{Caillo:05,Stell:95,Levin:96,Patsah:04}. Nevertheless, the critical
exponents we obtain from our SCOZA investigation for the RPM show a cross-over
behavior to MF.

The paper is organized as follows: after presenting our model (Sec.\ II), we
 give a brief introduction to the theoretical concepts that we apply to study
 the critical behavior of our system: an improved MF theory and SCOZA
 (Sec.\ III); the section is closed with a comparison between the data produced
 by these concepts. In Section IV we give details about the simulation
 techniques that  we apply and discuss the problems hereby encountered. In the
 subsequent section (Sec.\ V) we make a comparison between the data obtained in
 the theoretical approaches and in the simulations. The paper is closed with
 concluding remarks. An appendix provides further technical details about our
 reweighting procedure.

\section{Model and reduced units}

Using the RPM for $v_{\rm IS}(r)$, our potential $v(r)$ is actually
the interaction of a binary system; thus
\begin{equation}
v_{ij}(r) =  \left\{ \begin{array} {l@{~~~~~~~~}l}
                             \infty  & r \le \sigma \\
                \displaystyle  \frac{q_i q_j}{r} + \xi^2 v_{\rm MF}(r)
                                     & r > \sigma \\
                             \end{array}
                             \right. .
\label{potential2}
\end{equation}
where $v_{\rm MF}(r)$ is given in (\ref{2}). 

Numerical results will be expressed in reduced units (taking the
hard-sphere (HS) diameter $\sigma$ as unit of length): reduced
temperature $T^*=1/\beta^*$, where  $\beta^*
= q^2/k_{\rm B} T \sigma$ ($q = q_+ = - q_-$ charge of the spheres, $k_{\rm
B}$ Boltzmann constant, $T$ temperature), reduced configurational
chemical potential $\mu^*=\mu/k_{\rm B}T$, reduced volume $V^*=V/
\sigma^3$ and reduced 
density $\rho^* = N/V^*$($N$ the number of particles).
However, for notational convenience we will drop the stars throughout
the paper.   

\section{Theory}

\subsection{Improved mean field theory  (MF2L)}

The Kac potential energy $E_p(\mathcal{C})$ of a grand canonical (GC)
configuration $\mathcal{C}$, assuming, for instance, periodic boundary
conditions, is
\begin{equation}
E_p(\mathcal{C})
=\frac{1}{2V}\sum_{k\in\Lambda}\widetilde{v}_{\rm MF}(k)  
\widetilde{\rho}(k)\widetilde{\rho}(-k) \; ;
\end{equation}
where $\widetilde{v}_{\rm MF}(k)$ and $\widetilde{\rho}(k)$ are the
Fourier transforms of $v_{\rm MF}(r)$ and the microscopic density,
respectively.
For a fixed volume $V$, $\widetilde{v}_{\rm MF}(k)= \widetilde{\Phi}(k/\alpha)$ vanishes for all
non zero ($k\neq 0$) wavevectors of Fourier space in the limit $\alpha
\to 0$ and $\widetilde{v}_{\rm MF}(0)= \widetilde{\Phi}(0)$.  Thus
\begin{eqnarray}
E_p(\mathcal{C}) &=& \frac{1}{2V}
\widetilde{\rho}(0)^2  \widetilde{\Phi}(0) \nonumber \\ 
&=&-2 \pi q^2 V \rho^2 \; .
\end{eqnarray}
The GC partition function of
the model Eq.\ (\ref{potential}) is then obtained as \cite{Caillo:06}
\begin{eqnarray}
\Xi_{V}\left(\beta, \nu \right)&\simeq& \int_{0}^{\infty} d\rho \; \sigma^3 \;
\exp\left[ -V \left( - \nu \rho +\right.\right.\nonumber\\&&\left.\left.
+ \beta f_{\mathrm{RPM}}\left(
\rho\right) - 2 \pi \beta q^2 \xi^2\rho^2\right) \right] \; ,
\label{gcpf}
\end{eqnarray}
where we have replaced the sum over the number of particles by an
integral over the density which is valid for large systems and has no
consequences for the thermodynamic limit (here $\beta = 1/k_{\rm B}T$) .  In Eq.\ (\ref{gcpf}),
$f_{\mathrm{RPM}}(\rho,\beta)$ is the Helmholtz free energy per unit
volume of the RPM fluid at density $\rho$ and $\nu\equiv \beta \mu$.
Taking now the limit $V \to \infty$, one gets
\begin{eqnarray}
\label{lim}
&&\lim_{V \to \infty}-\frac{1}{V}\ln 
\Xi_{V}\left(\beta, \nu \right)=\min_{\rho}\mathcal{L}
 (\beta,\nu,\rho) \; ,\nonumber \\
&&\mathcal{L} (\beta,\nu,\rho)= -\nu \rho + \beta
f_{\mathrm{RPM}}(\rho,\beta) -2 \pi \beta q^2 \xi^2 \rho^2   \; .
\end{eqnarray}
The function $\mathcal{L} (\beta,\nu,\rho)$ plays the role of a Landau
function \cite{Caillo:06,GOLDEN}. Denoting $\beta_c^{\mathrm{RPM}}$ the
inverse critical temperature of the RPM, we note that $\beta
f_{\mathrm{RPM}}(\rho,\beta)$ is an analytical and {\it strictly} convex
function of $\rho$ for $\beta < \beta^{\mathrm{RPM}}_c$ and that the critical
point of our model occurs for $\beta <\beta^{\mathrm{RPM}}_c$. From the
analysis of the Kac model it can thus be inferred that the composite system
Eq.\ (\ref{potential}) will have MF behavior \cite{Kac:59,Caillo:06}.

Equation (\ref{lim}) can be taken as the starting point for an
improved mean field (MF) theory. To this end an approximate expression
for $f_{\mathrm{RPM}}(\beta,\rho)$ is inserted in the Landau function
$\mathcal{L} (\beta,\nu,\rho)$ of Eq.\ (\ref{lim}) and
$\mathcal{L}(\beta,\nu,\rho)$ minimized with respect to $\rho$ for
each inverse temperature $\beta$ and reduced chemical potential
$\nu$. At low temperatures and arbitrary $\nu$, $\mathcal{L}
(\beta,\nu,\rho)$ has in general two minima, $\rho_{g}$ and
$\rho_{l}$, but there is only one value of $\nu$, i.e.\ 
$\nu_{\mathrm{coex}}(\beta)$, for which the Landau function takes
equal values at the two minima. The corresponding minima are the
densities of the gas and liquid phases at coexistence, respectively.

We have used for $f_{\mathrm{RPM}}$ the loop-expansion of the RPM free
energy obtained in the field theoretical framework of Ref.\ 
 [\onlinecite{Caillo:06,Caillo:05}], i.e.
\begin{equation}
\label{loop}
f_{\mathrm{RPM}} = f_{\mathrm{RPM}}^{(0)} +
f_{\mathrm{RPM}}^{(1)} +f_{\mathrm{RPM}}^{(2)} + \ldots  
\end{equation}
where the superscripts $^{(p)}$ denote the $p^{\mathrm{th}}$ order
contribution to the loop expansion of $f_{\mathrm{RPM}}$.

At the  tree level one has for the RPM \cite{Caillo:05}
\begin{equation}
\label{f-0loop}
\beta f_{\mathrm{RPM}}^{(0)}= \beta f_{\text{HS}}(\rho) 
-\rho \ln 2 -\frac{\rho}{2}
\beta q^{2} v_c(0)\; ,
\end{equation}
where $f_{\text{HS}}(\rho)$ denotes the excess free energy per unit
volume of the reference HS fluid for which the Carnahan-Starling
approximation \cite{Carnah:69}  was used in numerical calculations.  The
$\rho \ln 2$
contribution stems from the entropy of mixing and $v_c(r)$ denotes the
Coulomb potential, $v_c(r)=1/r$ for $r > \sigma$ and regularized
inside the core region.  

The one-loop contribution $\beta f_{\mathrm{RPM}}^{(1)}$ reads
\begin{eqnarray}
\label{loop-1}
\beta f_{\mathrm{RPM}}^{(1)}=
 \frac{1}{2}\int \frac{d^{3}\mathbf{k}}{(2 \pi)^{3}}\; 
\ln\left(1+ \beta \rho q^{2} \widetilde{v}_c(k)\right) \; , 
\end{eqnarray}
and coincides with the free energy in the random phase approximation
(RPA) \cite{Caillo:05}. 

Finally, the two-loop contribution $\beta f_{\mathrm{RPM}}^{(2)}$,
still tractable, has expression
\begin{eqnarray} 
\label{f-2loop} 
 \beta f_{\mathrm{RPM}}^{(2)}=
- \frac{\beta^{2}}{4}
 [\rho q^{2}]^{2} \int  d^{3}\mathbf{r}\; 
 h_{\text{HS},\;\rho }(r) \Delta^{2}(r)    
 \; ,
\end{eqnarray} 
where $ h_{\text{HS},\;\rho }(r)$ denotes the usual pair distribution
function of hard spheres (HS) at number density $\rho $; $\Delta(r)$ the
propagator of the free theory whose expression in $k$- space is
\cite{Caillo:05} 
\begin{equation}
 \label{Deltak}
 \widetilde{\Delta}(k)=\frac{\widetilde{v}_c(k)}{1+\beta \;
 \rho q^{2} \;\widetilde{v}_c(k) } \ .
\end{equation}

As discussed at length in Ref.\ [\onlinecite{Caillo:05,Caillo:06}] the
loop-expansion is not a systematic expansion in some small physical
parameter and thus depends explicitely on the regularization of the
Coulomb potential $v_c(r)$ in the core ($r<\sigma$).  In this work we
adopted the MSA regularization \cite{Caillo:05} which ensures that, at
the one-loop level, the pair distribution function vanishes inside the
core. Moreover, with this specification the one-loop approximation of
$f_{\mathrm{RPM}}$ coincides with the optimized RPA (ORPA) theory or
the MSA theory if one adopts, as we did, the Percus-Yevick (PY)
expression \cite{HANSEN} for $h_{\text{HS},\;\rho }(r)$.

With the MSA regularization \cite{HANSEN}
 \begin{eqnarray}
 \label{MSA}
  v_{\text{c,MSA}}(r)&=& 1/r  \phantom{\frac{B}{\sigma} \; \left(2- \frac{B r }{\sigma}\right)}\quad  ( r\geq \sigma) \; , \\
  \label{jj}
  v_{\text{c,MSA}}(r)&=& \frac{B}{\sigma} \; \left(2- \frac{B r }{\sigma}\right)   
\phantom{ 1/r}\quad   ( 0 \leq r \leq \sigma)  \; ,               \\
  B&=& \frac{x^2 +x -x(1+2 x)^{1/2}}{x^2} \; ,
\end{eqnarray}
where $x=\kappa \sigma$ and $\kappa^2=4\pi\beta\rho q^2$ is  the
Debye wavenumber squared.

\subsection{SCOZA}

As noted in the introduction the formalism of SCOZA is not applicable
if we choose in our interaction $v_{ij}(r)$ $\alpha$ to be
zero. Therefore we consider a slightly modified potential, $\bar
v_{ij}(r)$, the so-called charged Yukawa model \cite{MolPhys_101_1611}
\begin{equation}
\bar v_{ij}(r) = v_{ij}^{\rm C}(r) + \xi^2 v^{\rm Y}(r) ~~~~~ i,j=+,-
\end{equation}
where $v_{ij}^{\rm C}(r)$ stands again for the RPM while the
Yukawa (Y) contribution reads
\begin{equation}
v^{\rm Y}(r) =  \left\{ \begin{array} {l@{~~~~~~~~}l}
                             \infty  & r \le \sigma \\
               -\displaystyle{ q^2 \frac{\alpha^{*2}}{r} {\rm e}^{-\alpha^* r/\sigma}}
                                     & r > \sigma \\
                             \end{array}
                             \right. .
\label{potential_Y}
\end{equation}
In the limit $\alpha^* \to 0$, $v^{\rm Y}(r)$ reduces to $v_{\rm
MF}(r)$.

For this particular system the set of OZ equations,
\begin{equation}
h_{ij}(r) = c_{ij}(r) + \sum_{k} \rho_k \left[ h_{ik} \otimes c_{kj} \right] (r) 
~~~~~~ i,j,k = +,- ,
\label{oz_bin}
\end{equation}
where '$\otimes$' denotes the convolution operation, can be decoupled:
with $\rho_- = \rho_+ = \rho/2$ and $q_+ = - q_- = q$ and introducing
the correlation functions
\begin{eqnarray} 
h^{\rm Y}(r) & = & \frac{1}{2} \left[ h_{++}(r) + h_{+-}(r) \right] 
\\
c^{\rm Y}(r) &=& \frac{1}{2} \left[ c_{++}(r) + c_{+-}(r) \right] \\
h^{\rm C}(r) & = & \frac{1}{2} \left[ h_{++}(r) - h_{+-}(r) \right] 
\\
c^{\rm C}(r) &=& \frac{1}{2} \left[ c_{++}(r) - c_{+-}(r) \right] 
\end{eqnarray}
the set of OZ equations (\ref{oz_bin}) decomposes into two
independently solvable one-component OZ equations for the Yukawa and
the Coulomb parts, namely
\begin{equation}\label{OZ_decompose}
\begin{array}{r@{\;=\;}l}
h^{\rm Y}(r) & c^{\rm Y}(r) + \rho \left[ h^{\rm Y} \otimes c^{\rm Y}\right](r) 
\\ 
h^{\rm C}(r) & c^{\rm C}(r) + \rho \left[ h^{\rm C} \otimes c^{\rm
    C}\right](r) .
\end{array}
\end{equation} 

The formalism of SCOZA \cite{MolPhys_95_483,JChemPhys_118_7414} is based on
the MSA closure to the OZ equations, i.e.,
\begin{equation}
\begin{array}{r@{\;=\;}l@{\quad}l}
g_{ij}(r) & 0  & {\rm for} ~~ r \le \sigma  \\ 
c_{ij}(r) & - \beta \bar v_{ij}(r)  & {\rm for} ~~ r > \sigma. 
\end{array}
\end{equation}
Introducing $g^{\rm Y}(r) = h^{\rm Y}(r) + 1$ and $g^{\rm C}(r) =
h^{\rm C}(r)$, leads to the MSA closure for the decoupled OZ equations
(\ref{OZ_decompose}):
\begin{equation}
\label{msa_Y}
\begin{array}{r@{\;=\;}l@{\quad}l}
h^{\rm Y}(r) & -1   &  r \le \sigma  \\ 
c^{\rm Y}(r) & c^{\rm HS}(r) - \beta v^{\rm Y}(r)  & r > \sigma 
\end{array}
\end{equation}
and 
\begin{equation}\label{msa_C}
\begin{array}{r@{\;=\;}l@{\quad}l}
h^\mathrm{C}(r) & 0 & r\le\sigma \\
c^\mathrm{C}(r) & -\beta\displaystyle \frac{q^2}{r} & r>\sigma \; .
\end{array}
\end{equation}
%
Above we have used for the correlation function of the HS reference
system, $c^{\rm HS}(r)$, the parameterization (for details see Ref.\ [\onlinecite{MolPhys_25_45}])
\begin{equation}
c^{\rm HS}(r) = K_1(\rho) \frac{\exp\left[-z_1(\rho)(r-\sigma)\right]}{r} .
\end{equation}

For the Yukawa part the formalism of SCOZA can be applied in a
straightforward way; introducing a yet unknown, state-dependent
function $K^{\rm Y}(\rho, T)$, the SCOZA closure relation (\ref{msa_Y})
reads
\begin{equation}
\label{scoza_Y}
\begin{array}{r@{\;=\;}l@{\quad}l}
h^{\rm Y}(r) & -1   & r \le \sigma \\
c^{\rm Y}(r) & c^{\rm HS}(r) - K^{\rm Y}(\rho, T) v^{\rm Y}(r)  & r > \sigma .
\end{array}
\end{equation}
For the Coulomb part the situation is more delicate: in Coulomb systems the
Stillinger-Lovett sum-rules \cite{HANSEN,JChemPhys_48_3858} have to be
satisfied which would be violated when introducing a function $K^{\rm C}(\rho,
T)$, similar to what we did for the Yukawa contribution in (\ref{scoza_Y}). We
therefore cannot extend SCOZA to the Coulomb part and have to treat it rather
within MSA, i.e., we use the closure relations (\ref{msa_C}) for $h^{\rm
C}(r)$ and $c^{\rm C}(r)$.

Within the SCOZA formalism a partial differential equation (PDE) is
derived that imposes thermodynamic self-consistency between the energy
and compressibility routes and thus fixes the yet undetermined
function $K^{\rm Y}(\rho, T)$. Thus we have to calculate the internal
energy and the compressibility of the system within the framework of
our theory. A straightforward analysis leads to
\begin{equation}
u = \frac{u^{\rm ex}}{V} = 2 \pi \sum_{i,j=+,-} \rho_i \rho_j \int dr r^2 g_{ij}(r)
\bar v_{ij}(r) = u^{\rm Y} + u^{\rm C}
\label{u_total}
\end{equation}
with $u^{\rm ex}$ being the excess (over HS) internal energy
and introducing
\begin{equation}
u^{\rm Y} = 2 \pi \rho^2 \int dr r^2 g^{\rm Y}(r) v^{\rm Y}(r)
\label{u_Y}
\end{equation}
and
\begin{eqnarray}
\beta u^\mathrm{C}\!\!\! &=&\! 2\pi\rho^2\beta \displaystyle
\int_0^\infty\!\!\!  g^\mathrm{C}(r)q^2 r\, dr=
\displaystyle\frac{w}{4\pi\sigma^3}\left(\sqrt{1+2w}-1-w \right) , \nonumber\\
w\!&=&\!\sqrt{{4\pi\rho\sigma^2q^2}/{k_\mathrm{B}T}} .
\end{eqnarray}
The explicit expression of $u^{\rm C}$  was derived
within MSA by Waisman and Lebowitz\cite{JChemPhys_56_3093}.

In a similar manner the reduced, dimensionless compressibility
$\chi_{\rm red} (= \rho k_{\rm B} T \chi_{\rm T})$ can be calculated
for our system, leading to
\begin{equation}
\left( \chi_{\rm red} \right)^{-1} = 1 - \frac{\rho}{2} 
\left[ \tilde c_{++}(0) + \tilde c_{+-}(0) \right] = 1 - \rho \tilde
c^{\rm Y}(0) 
\label{chi_red}
\end{equation}
where the tilde denotes the Fourier transform.

SCOZA imposes thermodynamic consistency between the energy and the
compressibility routes via the following PDE
\begin{equation}
\rho \frac{\partial^2 u}{\partial \rho^2} = 
\frac{\partial}{\partial \beta} \left(\frac{1}{\chi_{\rm red}} \right) .
\end{equation}
Using (\ref{u_total}) -- (\ref{chi_red}) and a few trivial
manipulations one ends up with the following equation
\begin{equation}
B(\rho, u^{\rm Y}) \frac{\partial u^{\rm Y}}{\partial \beta} + D(\rho, u^{\rm C}) =
C(\rho) \frac{\partial^2 u^{\rm Y}}{\partial \rho^2} ,
\label{scoza_pde}
\end{equation}
i.e., a PDE which has a structure similar to the standard SCOZA-PDE for the
one-component fluid (see e.g.\ Ref.\
[\onlinecite{EurophysLett_63_538}]). $B(\rho, u^{\rm Y})$ is formally
identical to the corresponding standard SCOZA-expression \cite{lisiPhD}
$B(\rho, u)$, and is obtained for the present problem by simply replacing $u$
by $u^{\rm Y}$; again, $C(\rho)=\rho$. Finally, one finds for the additional
coefficient $D(\rho, u^{\rm C})$ the following relation:
\begin{equation}
D(\rho, u^{\rm C}) = - \rho \frac{\partial^2 u^{\rm C}}{\partial \rho^2} .
\end{equation}
Once the PDE (\ref{scoza_pde}) is solved, further thermodynamic
quantities (notably the pressure and the chemical potential) follow
via standard thermodynamic relations \cite{EurophysLett_63_538,lisiPhD}.

While previous applications of SCOZA were rather dedicated to the
determination of the coexistence curves and to a reliable location
of the critical point within reasonable accuracy, we rather focus in
the present contribution on a {\it quantitative} description of the
critical properties. This means that we have to approach the critical
point very closely: introducing $\tau = (T - T_c)/T_c$, we need
accurate and reliable data down to $\tau \sim 10^{-10}$ or even
less. Compared to previous SCOZA applications this requires a
considerable increase in the numerical accuracy. We have realized this
goal by suitable modifications of the code and by increasing the
default numerical accuracy to four-fold (extended) precision. Taking these
measures is in particular indispensable for small $\alpha$- and
$\xi^2$-values where the Coulomb and Yukawa potentials at contact differ by
many orders of magnitude. Of course this increase in numerical
accuracy is accompanied by a considerable increase in computational
effort.

\subsection{Comparison between the theoretical approaches}

\begin{figure}[!htpb]
\includegraphics[clip=true, width=\columnwidth, draft=false]{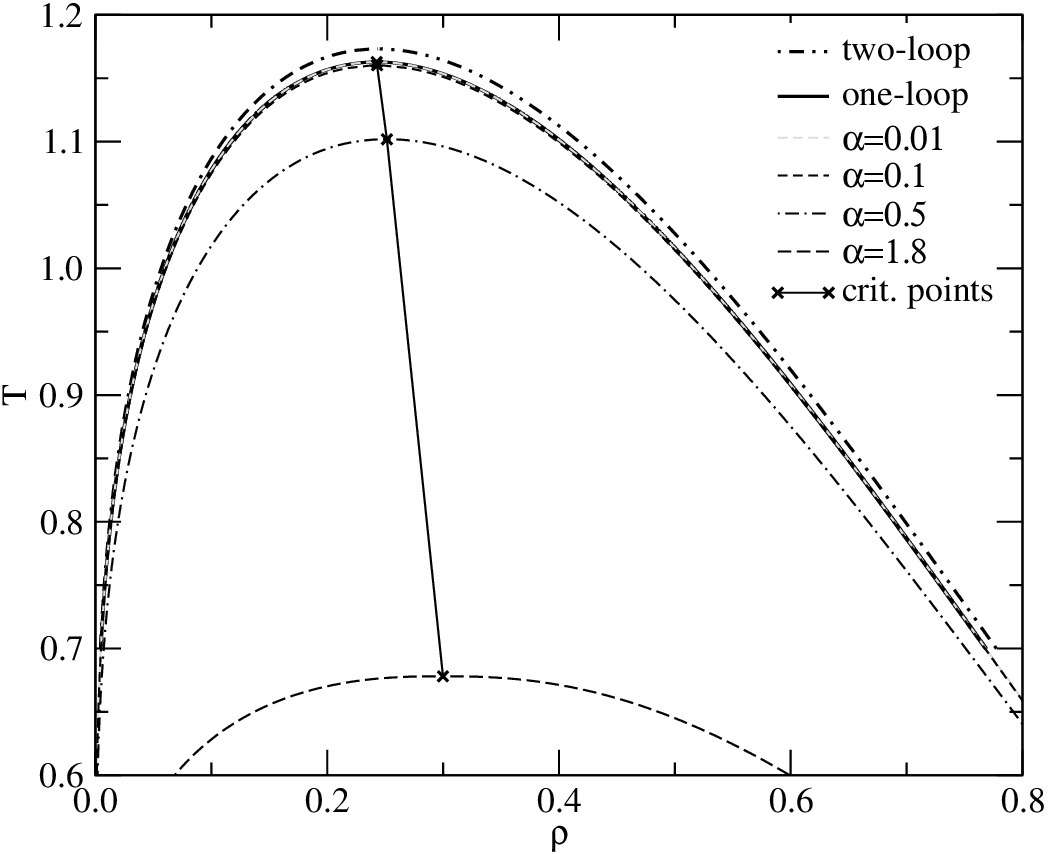}
\caption{Phase diagram in the $(T, \rho)$-plane of the system
investigated in the present study for $\xi^2 = 1$. Results were
obtained via the improved MF theory (both one- and two-loop level) and
SCOZA (for different values of $\alpha$ -- see text); lines as
indicated in the legend. In addition, a line connecting the respective
critical points is drawn.}
\label{fig1}
\end{figure}

In Figure \ref{fig1} we show the phase diagram of the system for $\xi^2 =
1$. We display results obtained from the improved MF theory both on the one-
and on the two-loop level. We show also SCOZA results, varying the index
$\alpha$ from 1.8 down to 0.01, which -- within numerical accuracy -- is
already very close to the Kac-limit. We observe that in this limit the SCOZA
data really 'converge' to the results obtained within the improved MF
theories. A more detailed analysis shows that for $\alpha = 0.01$ the SCOZA
data coincide with the results obtained on the one-loop level, while there is
a slight difference w.r.t.\ to the two-loop level.  This outcome is valid for
any value of $\xi$. Although a formal proof may be difficult, the limit
$\alpha \to 0$ of SCOZA will reproduce the exact MF result for $h^{\rm Y}(r)$ and
$c^{\rm Y}(r)$ while $h^{\rm C}(r)$ and $c^{\rm C}(r)$ are solution of the MSA. The
one-loop theory is precisely built with these ingredients and should coincide
with SCOZA as $\alpha \to 0$. The MF2L theory improves $h^{\rm C}(r)$ and $c^{\rm
C}(r)$ for all $\xi$ although for $\xi \to 0$ the improvement is minor (cf.\
Sec.\ V).  Comparison with simulation data will show that the two-loop level
indeed represents an improvement. An even more elaborate version of the
two-loop version of the improved MF theory that requires the pair distribution
function to vanish inside the core will certainly lead to an even better
agreement with simulations.

\begin{figure}[!htpb]
\includegraphics[clip=true, width=\columnwidth,draft=false]{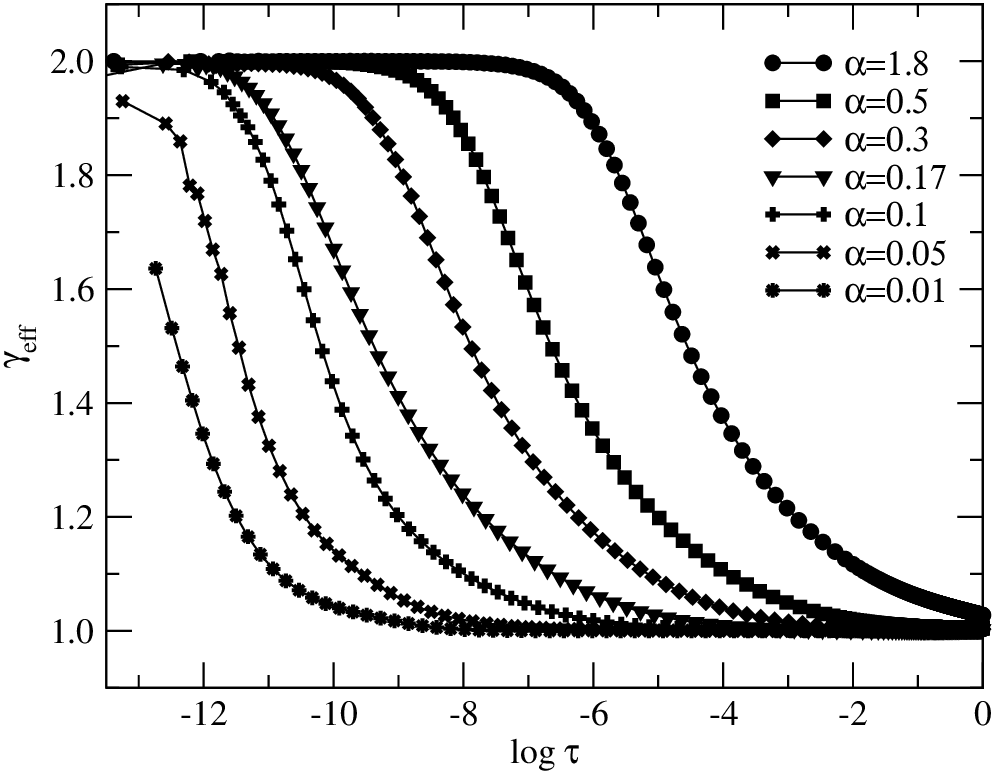}
\caption{(Effective) critical exponent $\gamma$ as a function of $\tau$
(for the definition see text) of the system investigated in the
present study for $\xi^2 = 0.16$ as obtained from SCOZA for different
values of $\alpha$, lined symbols as indicated in the legend.}
\label{fig2}
\end{figure}

In Figure \ref{fig2} we show results obtained for the effective critical
exponent $\gamma_\mathrm{eff}$ as a function of $\tau$, obtained from SCOZA
for $\xi^2 = 0.16$ and for a sequence of $\alpha$-values; $\alpha$ varies from
1.8 down to 0.01. $\gamma$ characterizes the divergence of the compressibility
as one approaches the critical point from {\it above}, i.e., $\chi_{\rm T}
\sim \tau^{-\gamma}$. The effective exponent $\gamma_\mathrm{eff}$ is obtained
by differentiating the logarithm of $\chi_\mathrm{T}$ w.r.t.\ the logarithm of
$\tau$ along the critical isochore. While in the MF universality class
$\gamma$ is 1, its SCOZA value\cite{Hoye:00} is 2.  Rather far from the
critical point (i.e., for $\tau \sim 10^{-2}$) we observe a MF behavior; as we
approach the critical point, a cross-over to the SCOZA behavior takes
place. In addition, the region where the SCOZA-value is attained shrinks
drastically as $\alpha$ is lowered. This reflects that in the Kac-limit
$\alpha \to 0$ the MF behavior becomes dominant, just as we expect it to
be. The curves presented in Figure \ref{fig2} indicate that for $\alpha = 0$
we would observe a MF behavior over the {\it entire} $\tau$-range.

\section{Simulations}

\subsection{Methods}

A detailed test of the validity of the theoretical approaches has been
made by comparison with GCMC simulations. 
The GCMC simulations are performed on a hypersphere using the approach,
detailed in Ref.\ [\onlinecite{Caillo:92,Caillo:93}], where the system of  
charged spheres is viewed as a single component fluid of charged bihard
spheres constrained to move at the surface of a four dimensional
sphere. The practical implementation of this method, which is
particularly adapted 
and efficient for simulating Coulomb systems of size $N< 10000$, can be
found in Ref.\ [\onlinecite{Caillo:97}]. 

Sampling of configuration
space of the system is made according to the GC probability
distribution
\begin{eqnarray}
&&p(N,R^N,\xi,\mu,T,V)=\\
&&={\frac{V^N \, \exp \left\{ \beta [ \mu N -U_{\rm RPM}(R^N)- U_{\rm MF}(N,V,\xi)]
\right\}}
{N! \, \Lambda^{3N} \Xi(T,V,\mu,\xi)}} \, .\nonumber
\label{proba}
\end{eqnarray}
Here $\Xi$ is the grand canonical partition function \cite{Caillo:93}  and $R^N$
indicates the positions  of the $N$ charged spheres (or charged bihard
spheres); 
$U_{\rm RPM}(R^N)$ is the energy of the RPM comprising Coulomb and
hard core interactions.  The term $U_{\rm MF}(N,V,\xi)=-a q^2 \xi^2 N^2$
(with $a=2\pi/V$) is the MF contribution to the internal energy as
explicited in Sec.\ IIIA.

For each value of $\xi$, we determine the critical point and, in its
vicinity, the coexistence line of the GL transition. The location of
the latter is obtained from the histograms 
%
\begin{eqnarray}
&&H(N,u,\xi,\mu,T,V) =\\
&&= C \int d\, R^N \, \delta 
\left[ u-U_{\rm RPM}(R^N)\right]\,  
p(N,R^N,\xi,\mu,T,V)\nonumber
\label{histo}
\end{eqnarray}
where $C$ is a constant of normalisation.  By integration of
$H(N,u,\xi,\mu,T,V)$ over $u$, one obtains the histogram
$h(N,\xi,\mu,T,V)$ which shows, in the vicinity of the GL coexistence
line two peaks located at $N=N_g$ and $N_l$, respectively,
and corresponding approximately to the densities $\rho_g$ and $\rho_l$ of
the gas and the liquid.

At given $T$ and $\xi$, the values of $\mu_e(L)$ ($L$ 
characteristic size of the system defined as $V^{1/3}$) corresponding to
GL equilibrium are affected by finite size effects. Therefore the
estimates of $\mu_e(L)$ are made for increasingly large volumes and
are then extrapolated to the thermodynamic limit.  Several procedures
are possible to define for a finite volume $V$ the states of GL
coexistence: First one can determine $\mu_e(L)$ such that the two peaks in
$h(N,\xi,\mu,T,V)$ have equal height or, preferably, equal area.  This
procedure, however, does not allow, at fixed $V$, a very precise
location of the critical point as in its vicinity the two peaks of
$h(N,\xi,\mu,T,V)$ tend to merge. A second method to locate, at fixed
$V$, phase equilibrium as well as critical temperatures and densities
has been proposed by Bruce and Wilding \cite{Wildin:92}. It relies on the use of
histograms of the variable ${\cal M}=\rho - s (u-aq^2 \xi^2 N^2)/V$
\begin{eqnarray}
&&p({\cal M},\xi,\mu,T,V)=\int du \sum_N \, 
\delta \left(\vphantom{\xi^2}{\cal M}- \rho +\right.\\\nonumber&&\qquad\left. + s \,(u-aq^2 \xi^2 N^2)/V \right) \, H(N,u,\xi,\mu,T,V) \, .
\label{ppmm}
\end{eqnarray}
For an appropriate choice of the
parameter $s$, ${\cal M}$ is such that at GL equilibrium and near the
critical point the distribution $p({\cal M},\xi,\mu,T,V)$ satisfies
the symmetry relation
\begin{equation} 
p\left( {\cal M} -\langle{\cal M}\rangle,\xi,\mu,T,V \right) = 
p\left( -{\cal M}+\left<{\cal M}\right>,\xi,\mu,T,V \right) 
\end{equation}
expected for the order parameter that characterizes phase transitions
of the Ising 3D or the MF universality class.
It is remarkable that, quantitatively, the histogram $h(N,\xi,\mu,T,V)$
corresponding to the symmetrized distribution  $p({\cal M},\xi,\mu,T,V)$
has two peaks of equal area.
A third possibility is to apply an unbiased finite size scaling method
presented by Fisher and co-workers \cite{Kim:03,Kim:05}.

If one defines $x=\left<\delta {\cal M}\right>/\left<\delta {\cal
M}^2\right>^{1/2}$ with $\delta {\cal M} = {\cal M} -\left<{\cal M}\right>$,
the critical temperature $T_c(L)$ is obtained by determining the values $\mu$,
$T$, and $s$ which enable to fit $p({x},\xi,\mu,T,V) \equiv p(x)$ to the
critical distribution $p_c({x})$ of the order parameter of a system of known
universality class. For example $p_c({x})$ can be the distribution of
magnetization of the Ising 3D model or that of the order parameter of a
lattice spin model of the mean field universality class.  For the Ising 3D
model $p_c({x})$ is known accurately from MC simulations \cite{Tsypin:00} and
exactly for the MF universality class (see below). The values of $T_c(L)$
obtained for increasingly large volumes $V$ can then be extrapolated to obtain
$T_c(\infty)$.


From the histograms $p({x})$ obtained at fixed volume the moments
$\left<\delta {\cal M}^2\right>$ and $\left<\delta {\cal M}^4\right>$ can be
calculated and from these the fourth order cumulant, $Q_L(\beta)=\left<\delta
{\cal M}^4\right>/\left<\delta {\cal M}^2\right>^2$, which in the vicinity of
$T_c(\infty)=1/\beta_c$ has the form \cite{Blote:95}
\begin{eqnarray} 
Q_L(\beta)\!\! &=& \!\! Q^* + a_1 \, (\beta -\beta_c )\,  L^{y_\tau}
+ a_2 \, (\beta -\beta_c )^2\,  L^{2y_\tau} +\nonumber \\&&\!\!+
a_3 \, (\beta -\beta_c )^3\,  L^{3y_\tau} + \ldots +
b_1 \,  L^{y_i} + \ldots \  . 
\label{qqfi}
\end{eqnarray}
$Q^*$ and the exponents $y_\tau$ and $y_i$ are characteristic of
the universality class of the system and can be determined, together
with $\beta_c$, by a fit of the $Q_L(\beta)$ obtained for different
system sizes.

For the MF universality class
$p(x)= A \exp{(-ax^4)}$ with $A= [\Gamma(3/4)]^2 2^{1/4} / \pi^{3/2}$, 
$a= \frac{1}{2} [\Gamma(3/4)]^4/ \pi^2$, $Q^*=2 [\Gamma(3/4)]^4 / \pi^2
= 0.456947$,  $y_\tau=d/2= 1.5$ and
$y_i=-d/2= -1.5$, respectively, where $d$ is the dimensionality of the
system \cite{LUIJTEN,Luijte:96}. As is well known the MF exponents do
not satisfy hyperscaling.
 
\subsection{Results}

Simulations have been performed for the values $\xi^2$ = 100, 11.11,
1.0, 0.1, 0.05, and 0.02, and volumes $V$ = 1000, 4000, 8000, and
16000. For each pair of $\xi^2$ and $V$ of the order of $5 - 10 \times
10^9$ MC trial moves were sampled.  At volume $V=16000$, even for
such a large number of configurations the statistical error remained
large so that the determination of phase equilibrium and the location
of the critical point was based solely on the smaller volumes.
Further, a reweighting procedure allowed to extend the results
obtained at the $\xi$-values listed above to neighboring values
of $\xi$ (see Appendix).
%

An approximate location of the GL transition inferred from the
appearance of two peaks in the histograms $h(N,\xi,\mu,T,V)$ indicates
that ($T_c, \rho_c$) varies from approximately (113, 0.25) at
$\xi^2=100$ to (0.075, 0.11) at $\xi^2=0.02$.  From these values we
can conclude that for the range of $\xi$ values considered the
critical temperature of our model is markedly higher than the critical
temperature of the RPM ($\xi^2=0$), i.e., $T_c \sim 0.049$ \cite{Caillo:02}.

It follows from the analyticity of the free energy (Sec.\ IIIA) that the
critical behavior of our model can be analyzed according to the classical
scheme (see e.g.\ Ref.\ [\onlinecite{LANDAU}]). In particular, the critical
exponents derived from the fit of $Q_L(\beta)$ should be MF-like and in the
procedure of Bruce and Wilding \cite{Wildin:92}, $p(x)$ has to be adjusted to
the distribution $p_c(x)$ of a model with MF criticality.
 
In Table \ref{tab1} we summarize the critical
temperatures $T^{Q}_c$ estimated from the fit of $Q_L(\beta)$
calculated for the volumes $V=$ 1000, 4000, and 8000 at the six
values of $\xi^2$ considered in the simulations.  These temperatures
are in good agreement with values for $T_c(L)$ obtained at $V$=8000 by fitting
$p({x})$ to the distribution $p_c(x)$ corresponding to the
MF universality class. In the table $\rho_c(L)$ denotes the average
density of the thermodynamic states at $\mu_c(L)$ and $T_c(L)$
corresponding to these fits.
  
The considered values of $\xi^2$ were too different to allow
reweighting in the range of  parameters  $\xi^2>0.1$.
 However, it was possible to
combine histograms obtained at $\xi^2$=0.05 and 0.1, on one hand, and
those at $\xi^2=$ 0.02 and 0.05, on the other. Furthermore, the
results at $\xi^2$ = 0.10 could be reweighted up to 0.13, and those at
$\xi^2=$ 0.02 to 0.01. However, combination of the
$H(N,u,\xi,\mu,T,V)$ was possible only at volume $V=$ 1000 due to
insufficient overlap of the energies at the larger volumes.
 
\begin{figure}[!htpb]
\includegraphics[clip=true,width=\columnwidth]{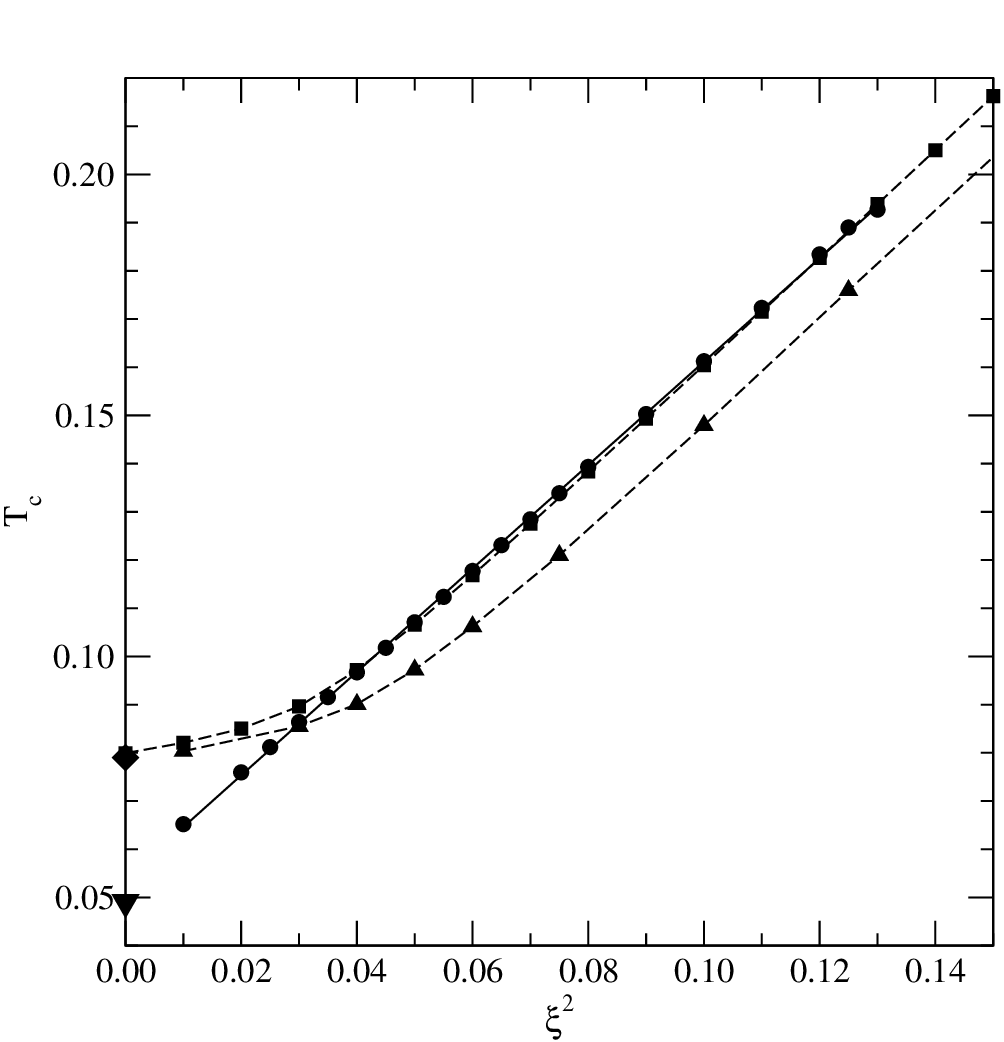} 
\caption{Variation of the critical temperature $T_c$ with strength $\xi^2$ of
  the Kac potential: MF2L -- squares and dashed line; SCOZA -- triangles up
  and dashed line; solid circles -- MC simulations for $V=1000$ (see
  text). The solid line is a linear fit of the simulation data. Diamond and
  triangle down indicate the critical temperatures of the RPM ( $\xi^2$=0)
  obtained by the MSA approximation \cite{Stell:95} and simulation
  \cite{Caillo:02,Panagi:02}, respectively.}
\label{fig3}
\end{figure}

\begin{figure}[!htpb]
\begin{center}
\includegraphics[clip=true,width=\columnwidth]{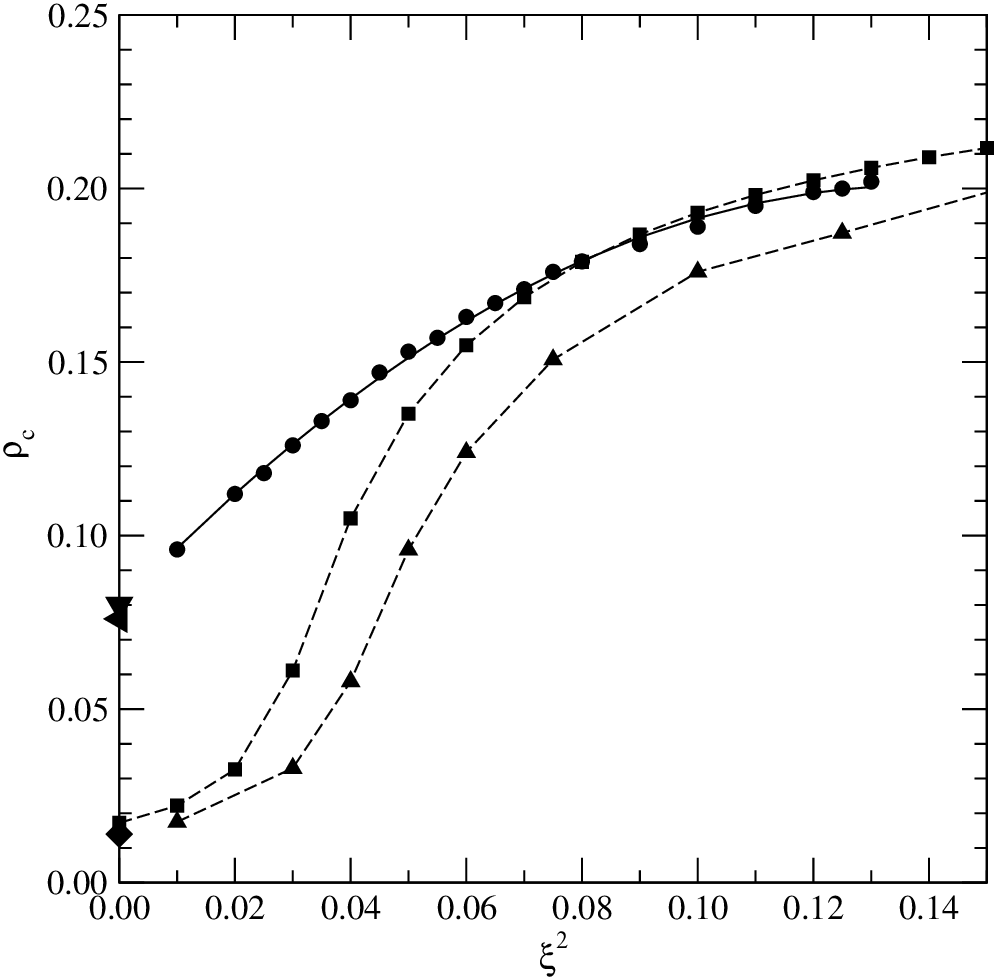} 
\caption{Same as Fig.\ 3 but for the critical density $\rho_c$. The MC results
  for the RPM are from Ref. \cite{Caillo:02} (triangles up) and
  Ref. \cite{Panagi:02} (triangles right), respectively.}
\label{fig4}
\end{center}
\end{figure}

\begin{figure}[!htpb]
\begin{center}
\includegraphics[clip=true,width=\columnwidth]{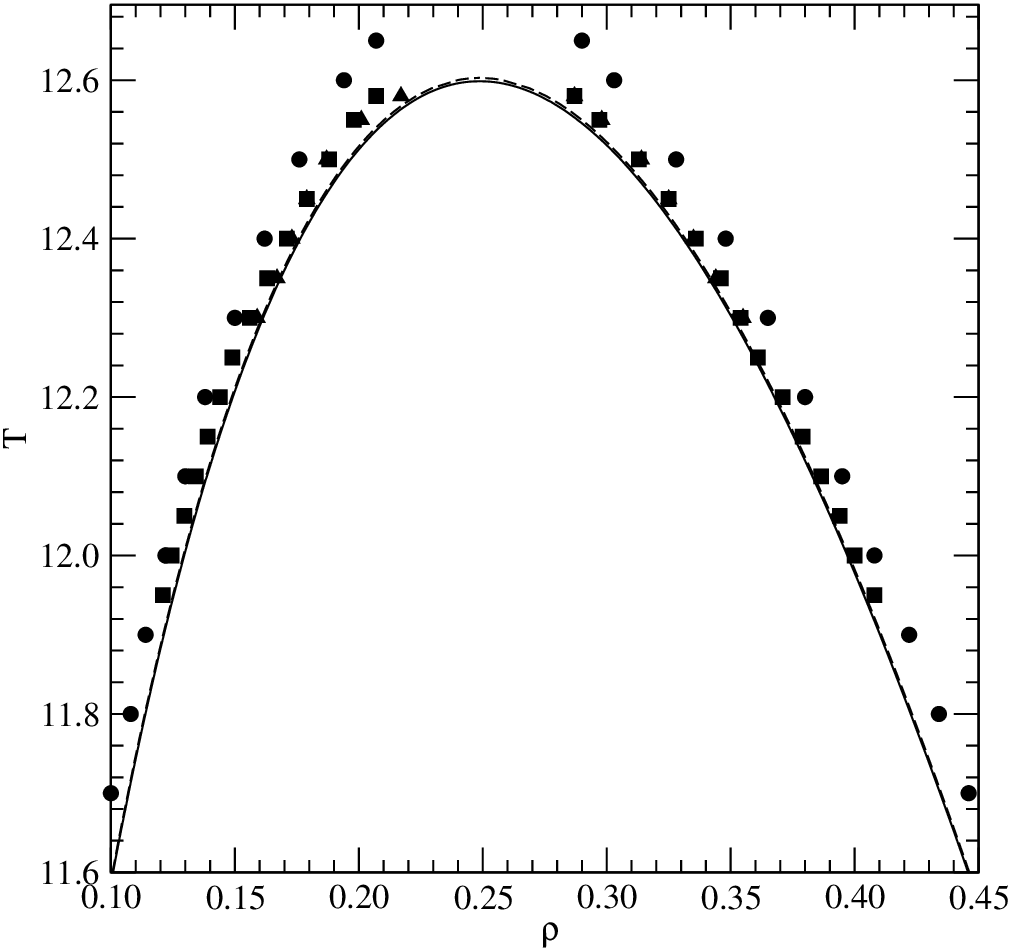} 
\caption{Comparison between simulation and theoretical estimates for the
  coexistence curve at $\xi^2 = 100/9$ in the vicinity of the critical
  point. The symbols represent MC results for the different volumes:
  $V=1000$ (circles), $V=4000$ (squares), $V=8000$ (triangles); MF2L (dashed
  line); SCOZA (full line).}
\label{fig5}
\end{center}
\end{figure}

\begin{figure}[!htpb]
\begin{center}
\includegraphics[clip=true,width=\columnwidth]{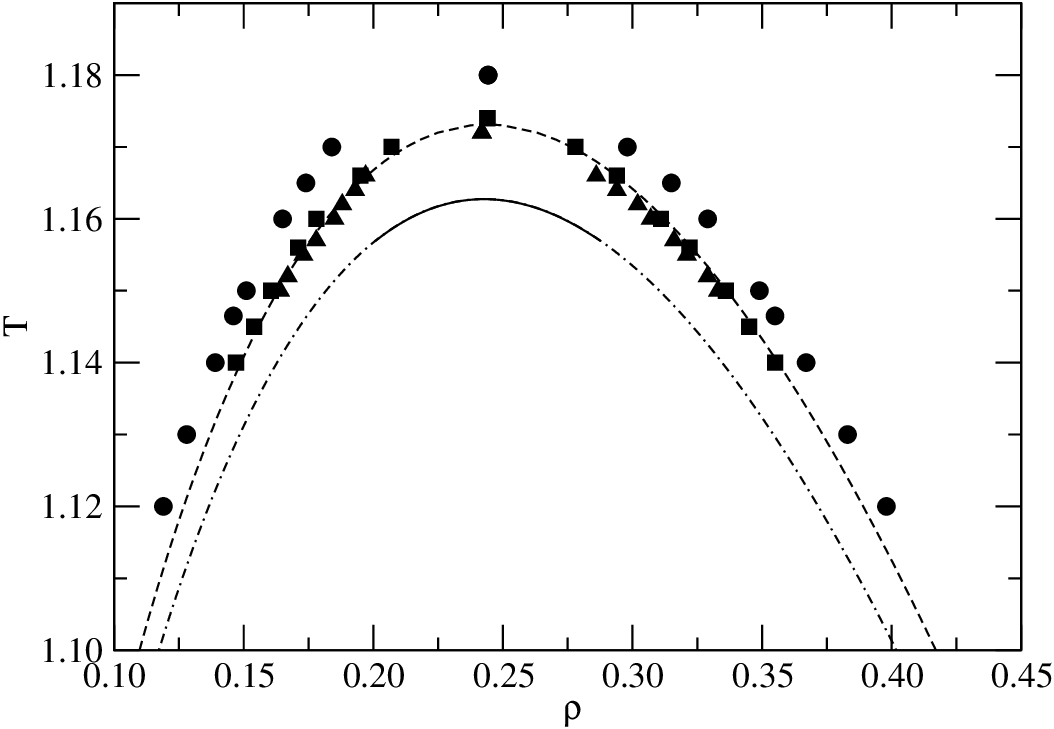} 
\caption{Coexistence curve at $\xi^2 = 1$ in the vicinity of the critical
  point. The symbols represent MC results for the different volumes:
  $V=1000$ (circles), $V=4000$ (squares), $V=8000$ (triangles); MF2L (dashed
  line); SCOZA (full line); one-loop approximation (dash-dotted line).}
\label{fig6}
\end{center}
\end{figure}

\begin{figure}[!htpb]
\begin{center}
\includegraphics[clip=true,width=\columnwidth]{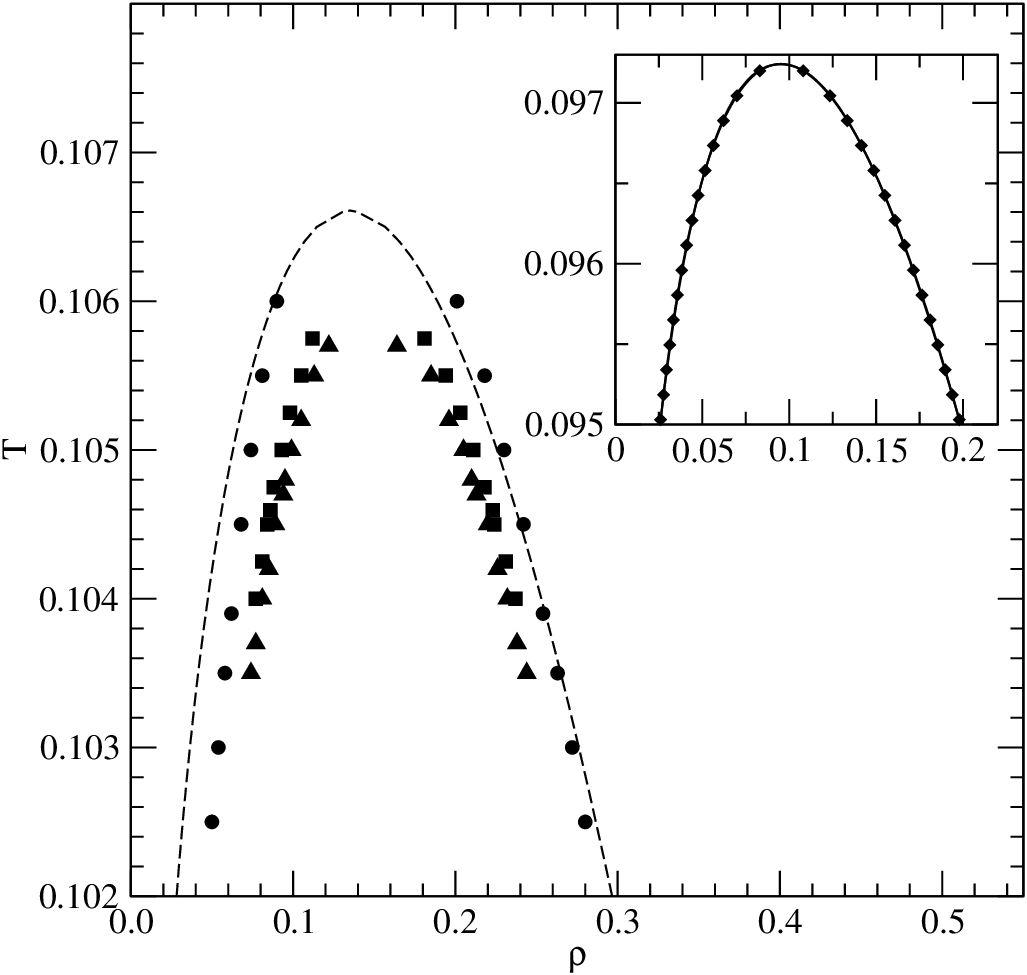} 
\caption{Coexistence curve at $\xi^2 = 0.05$ in the vicinity of the critical
  point. The symbols represent MC results for the different volumes:
  $V=1000$ (circles), $V=4000$ (squares), $V=8000$ (triangles); MF2L (dashed
  line). The inset shows SCOZA (full line) and the one-loop
  approximation (diamonds).}
\label{fig7}
\end{center}
\end{figure}

\begin{figure}[!htpb]
\begin{center}
\includegraphics[clip=true,width=\columnwidth]{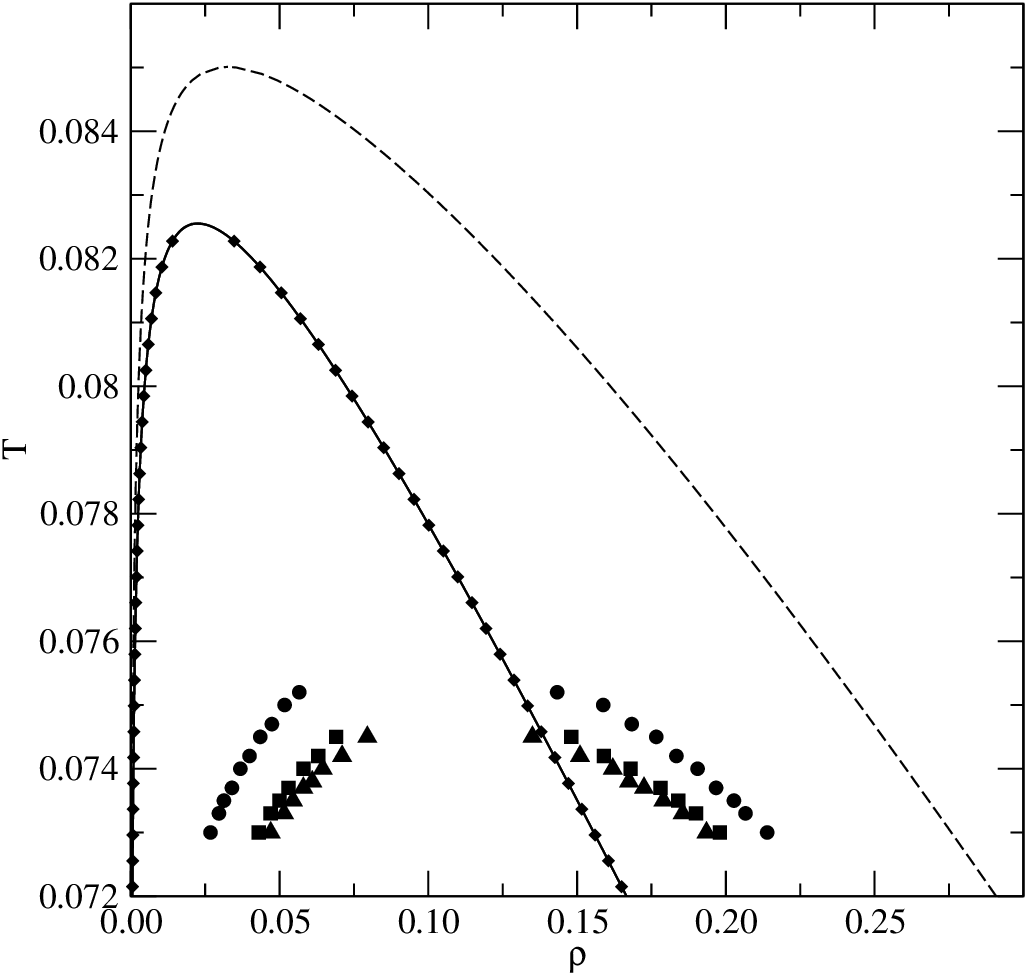} 
\caption{Coexistence curve at $\xi^2 = 0.02$ in the vicinity of the critical
  point. The symbols represent MC results for the different volumes:
  $V=1000$ (circles), $V=4000$ (squares), $V=8000$ (triangles); MF2L (dashed
  line); SCOZA (full line);  one-loop approximation (diamonds).}
\label{fig8}
\end{center}
\end{figure}

Figures \ref{fig3} and \ref{fig4} show the variation of $T_c(L)$
and $\rho_c(L)$ as a function of $\xi^2$ between $\xi$=0.01 and 0.13
for $V=1000$.  GL coexistence curves are presented in Figs.\
\ref{fig5} -- \ref{fig8} for $\xi^2$=0.02, 0.05, 1.0, and 
11.11  and the three volumes $V=$1000, 4000, and 8000.
The error in $T_c(L)$ is of the order of $\sim 0.3-0.1$\% depending
on the amount of sampling of the histograms $h(N,\xi,\mu,T,V)$.
$T_c(L)$ decreases weakly ($\sim 1$\%) when $V$ varies from 1000
to 8000 and differs from $T^Q_c$ by about $1\%$.  From these results
it is concluded that in the range 0.01$<\xi^2<$0.13 the values of
$T_c(L)$ obtained by reweighting for $V=$1000 provide a reliable
estimate of $T^Q_c$ or $T_c(L)$ at $V$=8000 with an uncertainty of
the order of $\sim 1\%$ (cf.\ Table II).

 The variation of the critical density with volume is small (cf.\ Table
 II), of the order 
of the statistical error on $h(N,\xi,\mu_c(L),T_c(L),V)$, estimated to
be $\sim 0.5-1.0$\% depending on sampling in the simulations. A second
estimate of $\rho_c$, denoted $\rho^d_c(L)$, is based on the rule of
rectilinear diameter, i.e., $\rho^d_c(L)=[\rho_g(T,L)+\rho_l(T,L)]/2$
valid at a MF critical point  for $T$ close to $T_c$.
The values of $\rho^d_c(L)$ given in Table II for
the different values of $\xi$ and $V$, were obtained by retaining only
the coexistence densities $\rho_g(T,L)$ and $\rho_l(T,L)$ which were
closest to the critical point at $T_c(L)$ and $\rho_c(L)$. The
coexistence densities $\rho_g(T,L)$ and $\rho_l(T,L)$ correspond to the
two maxima of $h(N,\xi,\mu,T,V)$ whose associated distribution $p({x})$
is symmetric (see above). The
uncertainty of the location of the maxima of $h(N,\xi,\mu,T,V)$
increases when $T$ approaches $T_c(L)$ so that the error on
$\rho^d_c(L)$ is of the order of $\sim \, 2$\%. The result for
$\rho^d_c(L)$ is somewhat lower than $\rho_c(L)$ though compatible
within the combined error estimates.
 
\begin{figure}[!htpb]
\begin{center}
\includegraphics[clip=true,width=\columnwidth]{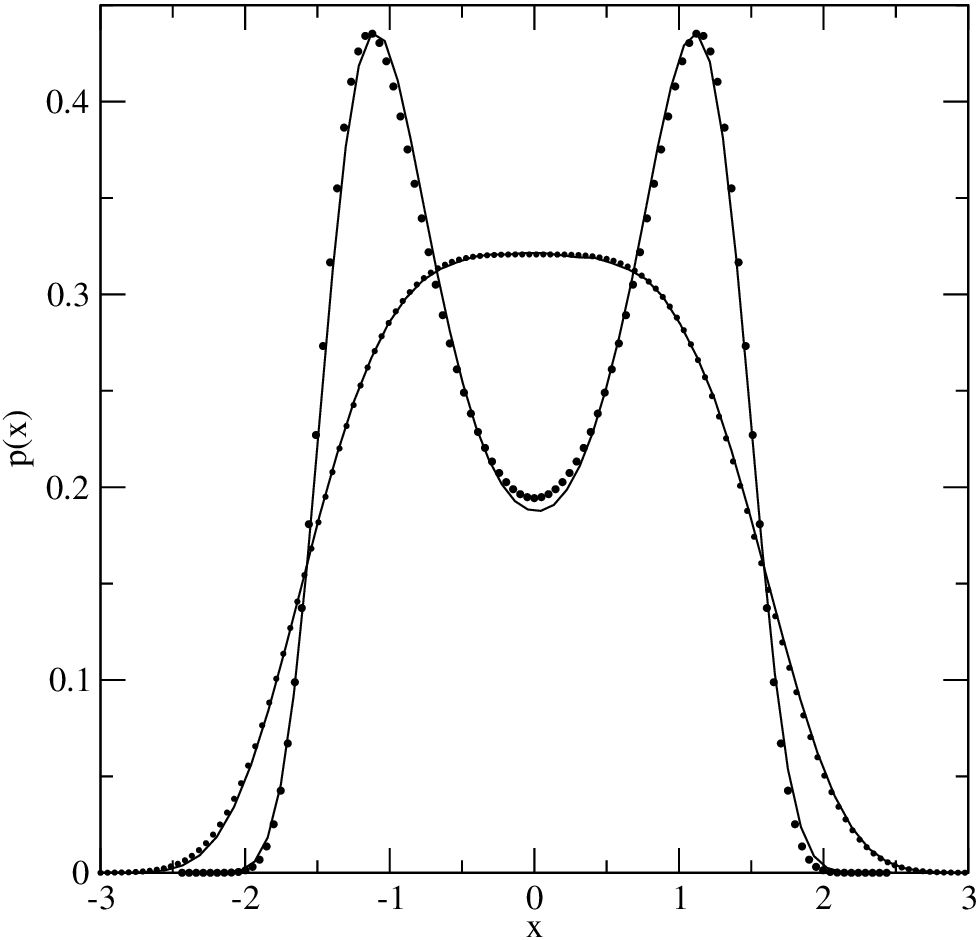} 
\caption{Matching of $p_c(x)$ Ising 3D ($T$=0.07385, $\mu$=-9.7953,
  $s=$ -1.03) (two peaks) and $p_c(x)$ MF ($T$=0.07477, $\mu$=-9.6901,
  $s=$ -0.98) at 
  $\xi^2 = 0.02$ and $V=$ 8000.}
\label{fig9}
\end{center}
\end{figure}

As mentioned previously in the literature \cite{Kim:04}, agreement of $p(x)$
with a $p_c(x)$ characteristic for a universality class, is necessary
but not sufficient to make unambiguous conclusions about the
universality class of a model system when studied by simulation.  This
uncertainty is illustrated for $\xi^2=0.02$ and $V=8000$ in Fig.\
\ref{fig9} which shows excellent fits of both the $p_c(x)$'s
correponding to the Ising 3D and MF universality classes.

\begin{figure}[!htpb]
\begin{center}
\includegraphics[clip=true,width=\columnwidth]{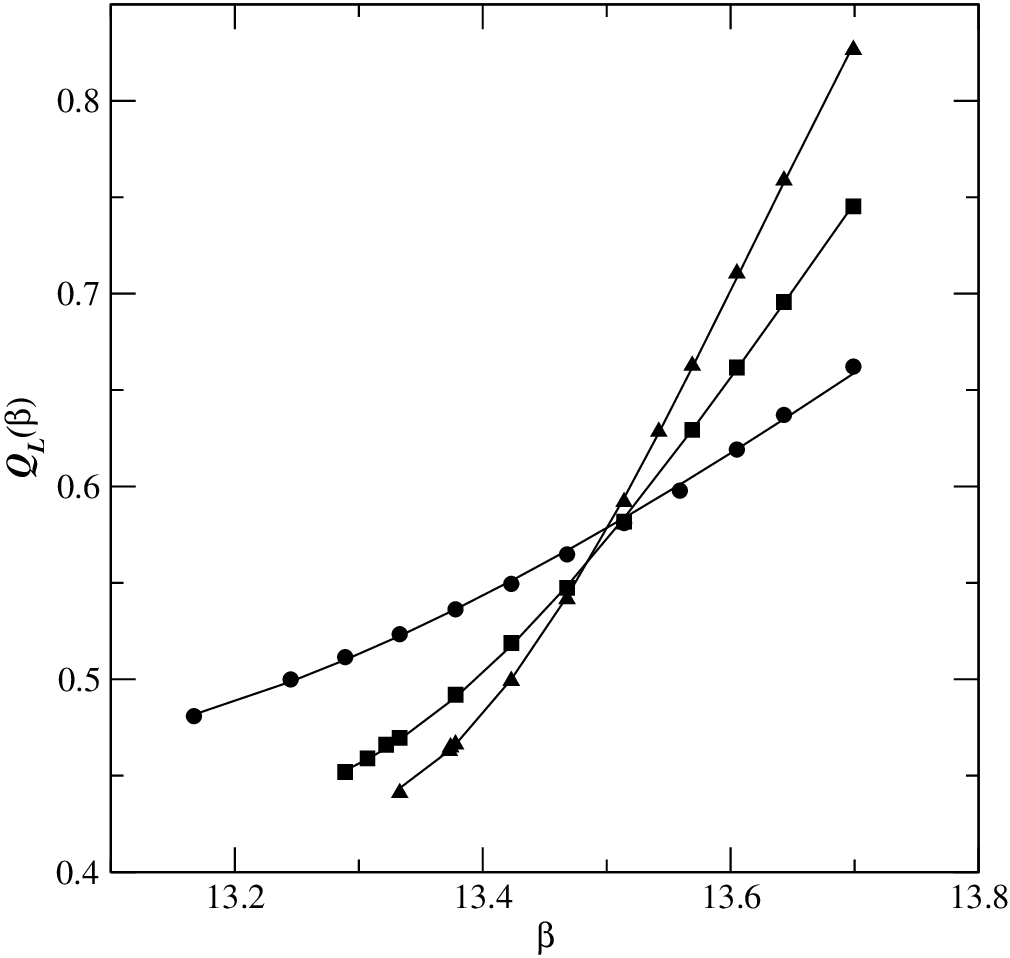} 
\caption{Fourth order cumulant $Q_L(\beta)$ as a function of inverse
 temperature $\beta$ at volumes $V=1000$ (solid circles), 4000 (squares),
 and 8000 (triangles) at $\xi^2 = 0.02$. The straight lines are fits of
 the simulation data to Eq.\ (\ref{qqfi}).}
\label{fig10}
\end{center}
\end{figure}

\begin{figure}[!htpb]
\begin{center}
\includegraphics[clip=true,width=\columnwidth]{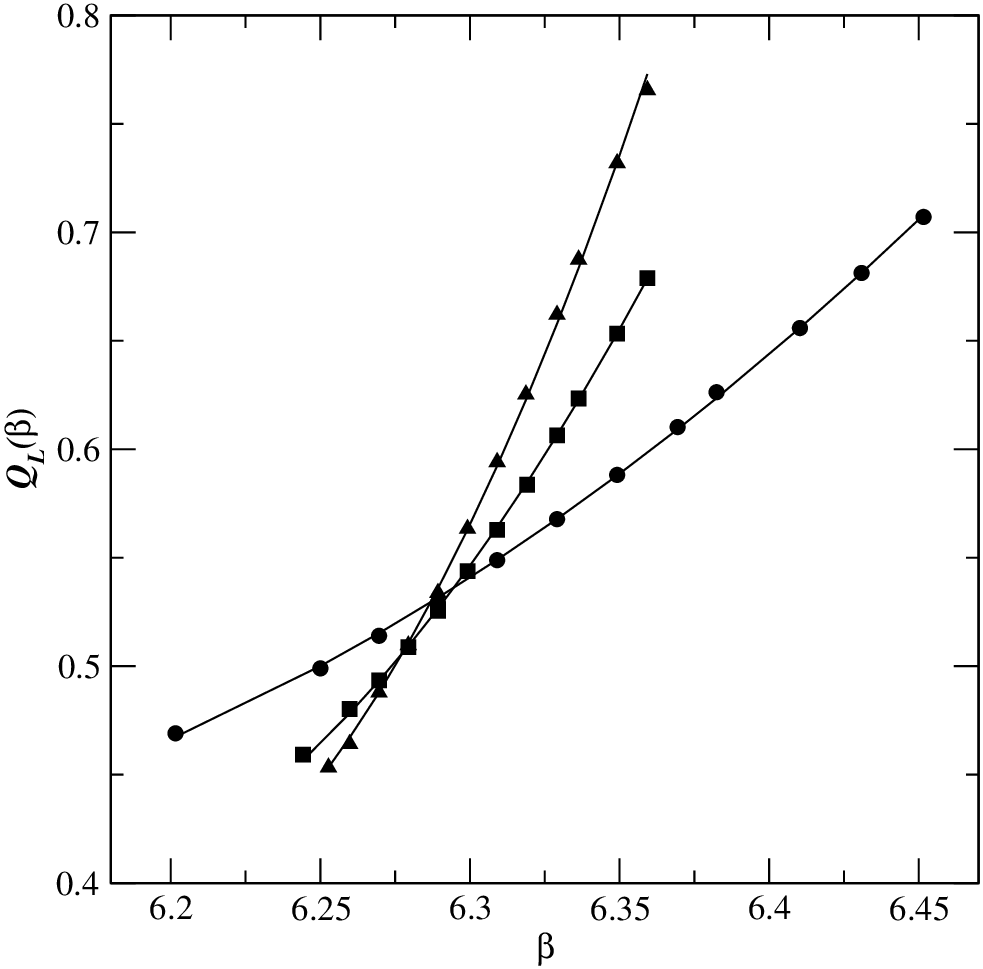} 
\caption{Fourth order cumulant $Q_L(\beta)$ as a function of inverse
 temperature $\beta$ at volumes $V=1000$ (solid circles), 4000 (squares),
 and 8000 (triangles) at $\xi^2 = 0.1$. The straight lines are fits of
 the simulation data to Eq.\ (\ref{qqfi}).}
\label{fig11}
\end{center}
\vspace{8cm}
\rule{0pt}{0pt}
\end{figure}

Although for the present model the MF universality class is implied by the
analyticity in $T$ and $\rho$ of the free energy $F(\xi, T, V)$ in the
vicinity of the critical points determined for $\xi^2>0.01$, this critical
behavior is also confirmed by the fit of the $Q_L(\beta)$ obtained for the
different values of $V$.  To realize the fits, $Q_L(\beta)$ was calculated,
for each volume, along the coexistence line by the procedure outlined above,
and, for $T>T_c(L)$, for values of $\mu$ and $s$ such that $p(x)$ remained
symmetric with respect to $x=0$.  For each volume, 8-12 values of $Q_L(\beta)$
were determined in the neighborhood of $T_c(L)$ [i.\ e.\ $\approx\, \pm 3$\%
from $T_c(L)$].  The fit is obtained by minimizing the square deviation
between the simulation results at the three volumes $V$ = 1000, 4000, and 8000
and Eq.\ (\ref{qqfi}) retaining in the expression the eight parameters shown
explicitly.  In Table III the values of $Q^*$, $\beta^Q_c$, $y_\tau$, and
$y_i$ are summarized for $\xi^2=$ 0.02, 0.05 and 0.1. The precision of the
fits can be appreciated from Figs.\ \ref{fig10} and \ref{fig11}. Taking into
account an uncertainty of $1-2$\% on $Q_L(\beta,L)$, fits compatible with this
error were obtained for a wide range of values of $Q^*$ and $y_i$, determined,
respectively, with a precision of $\sim$ 10 and $\sim$ 30\%.  In contrast, the
precision of $\beta^Q_c=1/T^Q_c$ is $\sim$ 1\% and of $y_\tau$ $\sim$ 5\%.
The value of $y_\tau$ is in agreement with the MF value and, despite large
uncertainties, also the values of $Q^*$ and $y_i$.

\section{Comparison: theories vs. computer simulations}   

While the results obtained of the two theories and the simulation
data have already been discussed in the respective subsections we
focus in the following on a direct comparison between the theoretical
and the simulation data.

To this end we go back to Table \ref{tab1} where
we have collected critical data (i.e., $T_c$ and $\rho_c$) for
different values of $\xi^2$ as obtained from the GCMC simulations, the
improved MF theory, and SCOZA; in the latter case we have chosen
$\alpha = 0.01$, i.e., a value that has turned out to be sufficiently
small to make the potential $v^{\rm Y}(r)$ Kac-like. From these data 
we can conclude that both theories are able to reproduce the critical
temperatures and critical densities for large and intermediate values
of $\xi^2 \gtrsim 1$ very accurately.
In the range of $\xi^2$ values $0.06-1$ the MF2L approximation
remains quantitative while SCOZA differs from the simulation results by
$\approx 10$\% for the critical temperature and $\approx 30$\% for the critical
density. Below $\xi^2 \sim 0.06$ 
the RPM contribution to the potential becomes dominant and we find,
not surprisingly, that the theoretical data start to differ from the
simulation results; in both theories, critical  temperature and density
tend to the MSA values of the RPM for $\xi^2 = 0$ whose failure to
reproduce the simulation data for the RPM
is well-documented \cite{Caillo:05}. As pointed out above SCOZA
identifies with the one-loop MF theory for all  $\xi^2$
as $\alpha \to 0$.

The coexistence curves in the vicinity of the critical point obtained
from theory and simulation   are   compared in
Figs.\ \ref{fig5} -- \ref{fig8} for selected values of $\xi^2$.
The convergence of the coexistence curves with system volume can be
appreciated from these figures; the results obtained for volume $V =
8000$ appear to be close to  the infinite system limit.
The MF2L approximation for the coexistence curve agrees well with simulation
data for $\xi^2 > 0.1$ but detoriates when approaching the RPM limit
where both the critical temperature and density are overestimated.
One should note, however, that even at  $\xi^2 = 0.02$ the critical
temperature differs from the MC result by only 10\%.
The MF2L approximation is a definite improvement over either the one-loop
approximation or SCOZA 
and could still be improved by requiring the pair distribution
function to vanish inside the core at the two-loop level.

As noted above, SCOZA provides data, that are in a less satisfactory agreement
with simulation data than those of the improved MF theory. Despite this
deficiency, SCOZA was able to provide useful information about the critical
behavior of the system: having non-classical critical exponents we have
been able to trace the cross-over from a MF to a non-MF behavior, as we
approach the Kac-limit (see Figure 2), verifying thus that the system
displays MF criticality for $\alpha = 0$.

\section{Conclusion and outlook}
We have investigated the critical behavior of a system where the
interaction is a linear combination of two potentials, each of them
having a well-defined critical behavior, {\it viz.} Ising 3D or
MF. Taking the RPM for the first contribution and a Kac-like potential
for the latter one, we could vary via a parameter $\xi^2$ the weight
of the respective contributions. Our investigations have been carried
out with GCMC simulations and two theoretical approaches, i.e., SCOZA
and the improved MF theory. The theoretical argument of Sec.\
III shows that  MF behavior at the critical point is expected for all 
$\xi^2 \ne 0$. Indeed  this
conclusion can unambiguously be drawn both from simulation and  SCOZA data.

We have given detailed account how the
simulations and the theoretical approaches perform in the critical
region. We could demonstrate that both simulations and theory have
reached a level of sophistication and accuracy that allows a {\it
quantitative} description of the critical behavior of simple liquids
in terms of the location of the critical point as well as of the
universality class. To be more specific, the location of the GL
critical point (based on the MF universality class) and the
coexistence lines are obtained from simulation within a numerical
accuracy of $\sim 1-2$\% with a slight variation with the volume $V$.
For our  model we find excellent
agreement between the theoretical approaches and computer simulations
for $\xi^2$-values down to $\sim 1$; the improved MF theory,
MF2L, performs in an excellent way even down to $\xi^2 \sim 0.05$. Only for
very small values of $\xi^2$, i.e., close to the pure RPM, simulation
data and theroretical results start to deviate. Nevertheless, SCOZA
predicts the correct critical behavior, i.e., MF.

We highlight the ambiguity arising when determining critical
temperatures
by fitting universal critical point order distribution functions by showing
that for the present system almost equally good fits are obtained for
both the universal Ising 3D and MF distributions.


Finally, by referring to the theoretical developments presented in Sec.\
IIIA we can remark that
the critical temperature is obtained by requiring that the second
derivative of $\mathcal{L} (\beta,\nu,\rho)$ with respect of the
density vanishes which yields
\begin{eqnarray}
\beta_c \partial^2 f_{\mathrm{RPM}}(\rho,\beta)/\partial \rho^2 \sim
\chi_T^{-1} = 2 \pi \beta_c q^2 \xi^2 \; ,
\end{eqnarray}
where $\chi_T$ denotes the isothermal compressibility of the RPM.  
Arguing that for small $\xi^2$,  $\chi_T \sim \vert T_c -T^{\mathrm{RPM}}_c
\vert^{\gamma}$ where $ \gamma $ is the critical exponent of the
compressibility of the RPM (or, more generally, that of the system with
potential $v_{\rm IS}$) it is  concluded that
\begin{eqnarray}
T_c -T^{\mathrm{RPM}}_c \sim \xi^{2/\gamma} \; .
\end{eqnarray}

This relation provides a means to evaluate the exponent $\gamma$. By
simulation this  turns out to be practical only for large systems ($V
\gtrsim 20000$) which are necessary 
to accurately sample
the density distribution function in the very low density  gas phase of the RPM
near the critical point.     
 
\begin{acknowledgements}

This work was supported by the \"Osterreichische Forschungsfond under
Project Nos. W004 and P17823-N08, and a grant through the Programme
d'Actions Int{\'e}gr{\'e}es AMADEUS under Project Nos.  06648PB and
7/2004, and by the Hochschuljubil\"aumsstiftung der Stadt Wien under Project
Number 1080/2002.
 
\end{acknowledgements}
   
\appendix*
\section{Reweighting}
 
Reweighting starts with the  histograms 
 $H_k(N,u,\mu_k,\beta_k,\xi_k) $ obtained with ${\cal N}_k$ 
 entries  at fixed volume $V$, chemical potential  $\mu_k$, and inverse temperature
 $\beta_k$ ($a=2 \pi /V$) 
\begin{eqnarray}
&&H_k(N,u,\mu_k,\beta_k,\xi_k) = 
{\frac{{\cal N}_k}{\Xi(\beta_k,\mu_k,\xi_k)}}  
\ \mathrm e^{\beta_k \, \mu_k N \, +\,\beta_k q^2  \xi^2_k a \, N^2}\cdot
\nonumber \\&&\quad \cdot \int dR^N \,
\delta\left(u-U_{{\rm RPM}}(R^N)\right) \, \mathrm e^{-\beta_k U_{{\rm
      RPM}}(R^N)} \nonumber =\\ 
 &&=  \Omega(N,u) \, {\frac{{\cal N}_k}{\Xi(\beta_k,\mu_k,\xi_k)}}  
\ \mathrm e^{\beta_k \, \mu_k N \, + 
\,\beta_k \, q^2 \, \xi^2_k a \, N^2 - \beta_k u}   
 \label{eq1} \end{eqnarray}

With $n$ histograms ($k=1, \, ... \, n$), 
an estimate of the density of states  $\Omega(N,u)$, up to a
multiplicative constant, is given by 
\begin{eqnarray}
\Omega(N,u) \ & \simeq  & {
\frac{ \sum_k \, H_k(N,u,\beta_k,\mu_k,\xi_k)} { \sum_k 
{\cal N}_k\ \mathrm e^{\beta_k \, \mu_k N \, + 
\,\beta_k \, q^2 \, \xi^2_k a \, N^2 - \beta_k u -F(\beta_k,\mu_k,\xi_k)}}}
 \nonumber \\& \simeq  & {
\frac{ \mathrm e^{F(\beta_1,\mu_1,\xi_1)}\sum_k \, H_k(N,u,\beta_k,\mu_k,\xi_k)} 
{ \sum_k 
{\cal N}_k\ \mathrm e^{\beta_k \, \mu_k N \, + 
\,\beta_k \, q^2 \, \xi^2_k a \, N^2 - \beta_k u -f(\beta_k,\mu_k,\xi_k)}}}
\nonumber
 \end{eqnarray} 
with $\mathrm e^{-F(\beta_k,\mu_k,\xi_k)}=\Xi(\beta_k,\mu_k,\xi_k)$ and
$\mathrm e^{-f(\beta_k,\mu_k,\xi_k)}=\mathrm e^{F(\beta_1,\mu_1,\xi_1)-F(\beta_k,\mu_k,\xi_k)}$.  

From the knowledge of $\Omega(N,u)$
any histogram 
$H_j(N,u,\xi_j,\mu_j,\beta_j,V) $ can then be calculated up to a
multiplicative constant, according to 
\begin{widetext}
\begin{equation}
H_j(N,u,\beta_j,\mu_j,\xi_j)  \simeq 
\frac{ 
 \mathrm e^{(\beta_j \mu_j\,- \, \beta_k\mu_k)  N \, + 
\, (\beta_j \,  \xi^2_j  \,-\beta_k \, q^2 \, \xi^2_k)\, a N^2 - 
(\beta_j \,- \, \beta_k)  u}  H_k(N,u,\beta_k,\mu_k,\xi_k)     } 
 {\int du \sum_N \,  
 \mathrm e^{(\beta_j \mu_j\,- \, \beta_k\mu_k)  N \, + 
\, (\beta_j \,  \xi^2_j  \,-\beta_k \, q^2 \, \xi^2_k)\, a N^2 - 
(\beta_j \,- \, \beta_k) u}  H_k(N,u,\beta_k,\mu_k,\xi_k)  } 
\end{equation}  

Using the initial guess $f_0(\beta_k,\mu_k,\xi_k)$
for $f(\beta_k,\mu_k,\xi_k)$, a first estimate of
 $\Omega(N,u)$ is obtained from 
\begin{equation}
{\bar H}(N,u,\beta_m,\mu_m,\xi_m) \, = \, { \frac{ \sum_k \, H_k(\beta_k,\mu_k,\xi_k,N,u)}
 {\sum_k 
{\cal N}_k\ \mathrm e^{(\beta_k \mu_k\,- \, \beta_m\mu_m)  N \, + 
\, (\beta_k \,  \xi^2_k  \,-\beta_m \, q^2 \, \xi^2_m)\, a N^2 - 
(\beta_k \,- \, \beta_m) \, u -f_0(\beta_k,\mu_k,\xi_k)}} }\nonumber
 \end{equation}
where the factor  $B_m(N,u)=\mathrm e^{\, - \, \beta_m \mu_m N  
\, - \, \beta_m \, q^2 \, \xi^2_m\, a N^2 \, - 
\, \beta_m \, u}$ is introduced to prevent too large  exponents
 in the sum   $\sum_k$ in the denominator with
 $\beta_m = (\sum_{k=1,..,n} \beta_k)/n$, 
 $\mu_m = (\sum_{k=1,..,n} \beta_k \mu_k)/(n \, \beta_m) $,
 and $\xi_m = (\sum_{k=1,..,n} \xi_k)/n$.
 
A new estimate of $f(\beta_k,\mu_k,\xi_k)$
is then obtained from 
\begin{equation}
\int du \sum_N 
{\bar H}(N,u,\beta_m,\mu_m,\xi_m) \, \,  \mathrm e^{(\beta_k \mu_k\,- \, \beta_m\mu_m)  N \, + 
\, (\beta_k \,  \xi^2_k  \,-\beta_m \, q^2 \, \xi^2_m)\, a N^2 - 
(\beta_k \,- \, \beta_m) \, u}=\mathrm e^{f_1(\beta_k,\mu_k,\xi_k)}
\end{equation}
where  $\int du$
is a discrete sum over the energy values $u$
effectively reached in the $n$ MC runs.
The values of  $f_1(\beta_k,\mu_k,\xi_k)$ lead to a new estimate 
of ${\bar H}$. Iteration is continued until for each
 $k$ we find
$|f_i(\beta_k,\mu_k,\xi_k)-f_{i-1}(\beta_k,\mu_k,\xi_k)|< 10^{-7}$.
After convergence ${\bar H}$ is normalized to 1.
In view of the factor
 $B_m(N,u)$ which relates ${\bar H}$ to $\Omega(N,u)$, ${\bar H}(N,u,\beta_m,\mu_m,\xi_m)$
is the histogram,  normalized to 1,  
corresponding to the average state  $\beta_m$, $\mu_m$, and $\xi_m$.

From  ${\bar H}$ one calculates the histograms 
 $H^{rew}_k(N,u,\mu_k,\beta_k,\xi_k) $ of states
 $\beta_k$, $\mu_k$, and $\xi_k$ by
 \begin{equation}
 H^\mathrm{rew}_k(N,u,\mu_k,\beta_k,\xi_k) = 
{\bar H}(N,u,\beta_m,\mu_m,\xi_m)  \, \, \mathrm e^{(\beta_k \mu_k\,- \, \beta_m\mu_m)  N \, + 
\, (\beta_k \,  \xi^2_k  \,-\beta_m \, q^2 \, \xi^2_m)\, a N^2 - 
(\beta_k \,- \, \beta_m) \, u}.
\label{eq3}
 \end{equation}
\end{widetext}
Comparison of the histograms calculated by MC with those obtained by the 
reweighting procedure provides a valuable test of the convergence of the MC
simulations.

\eject

\clearpage


\begin{table}[!htpb]
\begin{minipage}{0.7\textwidth}
\begin{center}
\vspace{1cm}
\caption{Variation of the critical temperatures and densities with
  $\xi^2$: $T_c^Q$ -- critical temperature obtained from the intersection of
  the cumulants $Q_L(\beta)$; $T_c(L)$  and $\rho_c(L)$ -- fit of
  simulation data at  $V$ = 8000 to the universal MF distribution;
  $T_c^S$  and $\rho_c^S$ -- SCOZA (with $\alpha = 0.01$);
  $T_c^{(2)}$  and $\rho_c^{(2)}$ -- MFL2. For statistical error see
  text. \\ }

\begin{tabular}{|c|c|c|c|c|c|c|c|}
\hline
\  \   \   $\xi^2$ \  \  \   & \  \    \  $T^Q_c$ \  \  \   
& \  \  \  $T_c(L)$ \  \  \   & \  \   \  $\rho_c(L)$ \  \   \  
 & \  \  \   $T^S_c$ \  \  \   & \  \   \   $\rho^S_c$ \  \  \  
 & \  \  \   $T^{(2)}_c$ \  \  \    
  & \  \  \    $\rho^{(2)}_c$ \  \  \     \\  \hline
   100.0  & 113.5   &  113.4 &  0.250   & 113.2 &  0.249 &  113.2  &  0.249  \\ \hline
   100/9  &   -     &  12.62 &  0.251   &  12.60 &  0.249 &  12.60 &  0.249  \\ \hline
    1.0   &   -     &  1.172 &  0.242   &  1.163 &  0.243 &  1.173  &  0.243  \\ \hline
    0.10  & 0.1596  & 0.1599 &  0.190   & 0.148  & 0.176  & 0.1604  &  0.193  \\ \hline
    0.05  & 0.1054  & 0.1062 &  0.151   & 0.0972 & 0.0959 & 0.1066  &  0.135  \\ \hline
    0.02  & 0.0742  & 0.0748 &  0.111   & 0.0825 & 0.0225 & 0.0850  &  0.032  \\ \hline
\end{tabular}
\end{center}
\end{minipage}
\label{tab1}
\end{table}


\begin{table}[!htpb]
\begin{minipage}{0.7\textwidth}
\begin{center}
\vspace{1cm}
\caption{Variation of the critical temperatures and densities with
  system size and  $\xi^2$: $T_c(L)$  and $\rho_c(L)$ -- fit of simulation
  data at volume  $V$ to the universal MF distribution;  $\rho_c^d(L)$ --
  from rule of rectilinear diameter (see text). For statistical error
  see text. \\ } 
\begin{tabular}{|c|c|c|c|c|}
\hline
\  \  \  \  $\xi^2$ \  \  \  \  &  \  \  \  \  $V$\  \  \  \  &  \  \ 
\  \  \  \  $T_c(L)$ \  \  \  \  & \  \  \  \  $\rho_c(L)$ \  \  \  \   & 
\  \  \  \  $\rho^d_c(L)$ \  \  \  \   \\  \hline 
 0.020 & 1000	&  0.07595  & 0.112 & 0.102 \\  
 0.020 & 4000	&  0.07506  & 0.111 & 0.108 \\ 
 0.020 & 8000	&  0.07477  & 0.111 & 0.108 \\  \hline 
 0.050 & 1000	&  0.10709  & 0.153 & 0.146 \\  
 0.050 & 4000	&  0.10620  & 0.151 & 0.147 \\ 
 0.050 & 8000	&  0.10625  & 0.151 & 0.143 \\  \hline 
 0.100 & 1000	&  0.16125  & 0.189 & 0.183 \\  
 0.100 & 4000	&  0.16015  & 0.189 & 0.186 \\ 
 0.100 & 8000	&  0.15993  & 0.190 & 0.187 \\  \hline 
 1.0 & 1000   &  1.181    & 0.244 & 0.244 \\  
 1.0 & 4000   &  1.174    & 0.244 & 0.243 \\ 
 1.0 & 8000   &  1.172    & 0.242 & 0.239 \\  \hline 
 100.0 & 1000	&  114.0    & 0.250 & 0.249 \\  
 100.0 & 4000	&  113.5    & 0.245 & 0.248 \\ 
 100.0 & 8000	&  113.4    & 0.247 & 0.246 \\  \hline 
\end{tabular}
\label{tab2}
\end{center}
\end{minipage}
\end{table}

\eject

\begin{table}[!htpb]
\begin{minipage}{0.7\textwidth}
\begin{center}
\vspace{1cm}
\caption{Values of critical indices $Q^*$, $y_{\tau}$, and $y_i$ obtained
  by the fit of $Q_L(\beta)$ [cf.\ Eq.\ (\ref{qqfi})] for
different values of $\xi^2$. The fits have been performed
with eight parameters. \\ }
\begin{tabular}{|c|c|c|c|c|}
\hline
\  \  \  \  $\xi^2$ \  \  \  \   & \  \  \  \   $Q^*$  \  \  \  \   & \  \  \  \   $\beta^Q_c$ \  \  \  \   
& \  \  \  \  $y_\tau$ \  \  \  \   & \  \   \  \  $y_i$ \  \   \  \    \\  \hline
0.02  & 0.53$\pm$0.05   &  13.47$\pm$0.1    &   1.54$\pm$0.1 & -1.49$\pm$0.5   \\  \hline
0.05  & 0.51$\pm$0.05   &  9.48$\pm$0.1    &    1.56$\pm$0.1 & -1.49$\pm$0.5   \\  \hline
0.10  & 0.47$\pm$0.05   &  6.27$\pm$0.1    &    1.45$\pm$0.1 & -1.57$\pm$0.5   \\  \hline
\end{tabular}
\label{tab3}
\end{center}
\end{minipage}
\end{table}



\begin{widetext}
\printtables
\end{widetext}

\clearpage

\printfigures


\begin{thebibliography}{99}

\bibitem{Caccam:96} C. Caccamo,
Phys.\ Rep.\ {\bf 274}, 1 (1996).

\bibitem{Parola:95} A. Parola and L. Reatto,  
Adv.\ Phys.\ {\bf 44}, 211 (1995).

\bibitem{Parola:84} A. Parola and L. Reatto,  
Phys.\ Rev.\ Lett.\  {\bf 53}, 2417 (1984);
Phys.\ Rev.\ A  {\bf 31}, 3309 (1985); for a more recent overview see also
A. Reiner and G. Kahl, Phys.\ Rev.\ E {\bf 65}, 046701-1 (2002);
A. Reiner and G. Kahl, J. Chem. Phys. {\bf 117}, 4925 (2002);
D. Pini, M. Tau, A. Parola, and L. Reatto, Phys.\ Rev.\ E {\bf 67}, 046116-1 (2003).

\bibitem{HANSEN} J.-P. Hansen and I. McDonald, {\it Theory of Simple Liquids},
 (Cambridge Univ. Press, Cambridge, 2006), 3rd edition.

\bibitem{Stell:69} G. Stell,   
Phys.\ Rev.\   {\bf 184}, 135 (1969).

\bibitem{Wildin:92} N.B. Wilding and A.D. Bruce,
J.\ Phys.:\ Condens.\ Matter {\bf 4}, 3087 (1992).

\bibitem{Kim:04} Y.C. Kim and M.E. Fisher,
J.\ Phys.\ Chem.\ B {\bf 108}, 6750 (2004).

\bibitem{Kim:03} Y.C. Kim, M.E. Fisher, and E. Luijten,
Phys.\ Rev.\ Lett.\ {\bf 91}, 065701 (2003).

\bibitem{Kac:59} M. Kac,
Phys.\ Fluids\  {\bf 2}, 8 (1959).

\bibitem{Lebowi:66} J.L. Lebowitz and O. Penrose,
J.\ Math.\ Phys.\ {\bf 7}, 98 (1966).

\bibitem{Hoye:77} J.S. H{\o}ye and G. Stell,  
J.\ Chem.\ Phys.\  {\bf 67}, 439 (1977).

\bibitem{Hoye:84} J.S. H{\o}ye and G. Stell,  
Mol.\ Phys.\  {\bf 52}, 1071 (1984).

\bibitem{MolPhys_95_483}
D. Pini, G. Stell, and N.~B. Wilding, Mol.\ Phys.\ {\bf 95}, 483 (1998).

\bibitem{MonatshChem_132_1413}
G. Kahl, E. Sch{\"o}ll-Paschinger, and A. Lang, Monatsh.\ Chem.\ {\bf 132},
1413 (2001).

\bibitem{EurophysLett_63_538}
E. Sch{\"o}ll-Paschinger and G. Kahl, Europhys.\ Lett.\ {\bf 63},  538  (2003).

\bibitem{Hoye:00} J.S. H{\o}ye, D. Pini, and G. Stell,  
Physica A  {\bf 279}, 213 (2000).

\bibitem{Caillo:05} J.M. Caillol,
Mol.\ Phys.\, {\bf103}, 1271 (2005); arXiv:cond-mat/0409455.

\bibitem{Caillo:06} J.-M. Caillol,
Mol.\ Phys.\ {\bf 104}, 1931 (2006);  arXiv:cond-mat/0602205.

\bibitem{Caillo:92} J.M. Caillol and D. Levesque,
J.\ Chem.\ Phys.\ {\bf 94}, 597 (1992).

\bibitem{Caillo:93} J.M. Caillol,
J.\ Chem.\ Phys.\ {\bf 99}, 8953 (1993).

\bibitem{Swends:93} R.H. Swendsen,
Physica A  {\bf 194}, 53 (1993).

\bibitem{Wildin:95} N.B. Wilding,
Phys.\ Rev.\ E {\bf 52}, 602 (1995).

\bibitem{Caillo:02} J.M. Caillol, D. Levesque, and J.-J. Weis,
J.\ Chem.\ Phys.\  {\bf 116}, 10794 (2002).

\bibitem{Luijte:02} E. Luijten, M.E. Fisher, and A.Z. Panagiotopoulos
Phys.\ Rev.\ Lett.\ {\bf 88}, 185701 (2002).

\bibitem{Panagi:02} A. Panagiotopulos,
J.\ Chem.\ Phys.\  {\bf 116}, 3007 (2002).

\bibitem{Stell:95} G. Stell,
J.\ Stat.\ Phys.\, {\bf78}, 197 (1995).

\bibitem{Levin:96} Y. Levin and M.E. Fisher,
Physica A  {\bf 225}, 164 (1996).

\bibitem{Patsah:04} O.V. Patsahan,
Condens.\ Matter Phys.\ {\bf 7}, 35 (2004).

\bibitem{GOLDEN} N. Goldenfeld, {\it Lectures on Phase Transitions and
 the Renormalization Group},
 (Addison-Wesley, New-York, 1992).

\bibitem{Carnah:69} N.F. Carnahan and Starling,  
J.\ Chem.\ Phys.\  {\bf 51}, 635 (1969).

\bibitem{MolPhys_101_1611}
T. Krist\'of, D. Boda, J. Liszi, D. Henderson, and E. Carlson, Mol.\ Phys.\ {\bf
  101},  1611  (2003).

\bibitem{JChemPhys_118_7414}
E. Sch{\"o}ll-Paschinger and G. Kahl, J.\ Chem.\ Phys.\ {\bf 118}, 7414 (2003).

\bibitem{MolPhys_25_45}
E. Waisman, Mol.\ Phys.\ {\bf 25},  45  (1973).

\bibitem{JChemPhys_48_3858}
F.~H. Stillinger and R. Lovett, J.\ Chem.\ Phys.\ {\bf 48},  3858  (1968).

\bibitem{JChemPhys_56_3093}
E. Waisman and J.~L. Lebowitz, J.\ Chem.\ Phys.\ {\bf 56},  3093  (1972).

\bibitem{lisiPhD}
E. Sch{\"o}ll-Paschinger, Ph.D.\ thesis, Institut f{\"u}r Theoretische Physik,
  TU Wien, Wiedner Hauptstr.\ 8-10, A-1040 Wien, Austria, 2002.

\bibitem{Caillo:97} J.M. Caillol, D. Levesque, and J.-J. Weis,
J.\ Chem.\ Phys.\  {\bf 107}, 1565 (1997).

\bibitem{Kim:05} Y.C. Kim and M.E. Fisher,
Comput.\ Phys.\ Commun.\ {\bf 169}, 295 (2005).

\bibitem{Tsypin:00} M.M. Tsypin and H.W.J. Bl\"ote,
Phys.\ Rev.\ E {\bf 62}, 73 (2000).

\bibitem{Blote:95} H.W.J. Bl\"ote, E. Luijten, and J.R. Heringa,
J.\ Phys.\ A: Math.\ Gen.  {\bf 28}, 6289 (1995).

\bibitem{LUIJTEN}  E. Luijten,
{\em Interaction Range, Universality and Upper Critical Dimension}
(Delft University Press, Delft, 1997).

\bibitem{Luijte:96} E. Luijten and H.W.J.  Bl\"ote,
Phys.\ Rev.\ Lett.\ {\bf 76}, 1557 (1996).

\bibitem{LANDAU} L.D. Landau and E.M. Lifshitz, {\it Statistical Physics},
 (Pergamon Press, London, 1959).






























\end{thebibliography}
\end{document}